\documentclass[longauth]{aa}
\usepackage{graphicx}
\usepackage{txfonts}
\usepackage[colorlinks, citecolor=blue, linkcolor=blue]{hyperref}
\usepackage{adjustbox}
\usepackage{xcolor}
\usepackage{float}
\usepackage{tabularx}
\usepackage{threeparttable}
\usepackage{hyperref}
\usepackage{siunitx} 
\newcommand{\sherlock}{{\fontfamily{pcr}\selectfont SHERLOCK}\,}
\newcommand{\matrixtk}{{\fontfamily{pcr}\selectfont MATRIX}\,}
\DeclareSIUnit\pixel{pix}
\DeclareSIUnit\year{yr}
\DeclareSIUnit\mag{mag}
\DeclareSIUnit\pc{pc}
\DeclareSIUnit\mas{mas}
\let\orgautoref\autoref
\renewcommand{\autoref}
        {\def\equationautorefname{Eq.}%
         \def\figureautorefname{Fig.}%
         \def\sectionautorefname{Sect.}%
         \def\subsectionautorefname{Sect.}%
         \def\subsubsectionautorefname{Sect.}%
         \orgautoref}

\makeatletter
\renewcommand*\aa@pageof{, page \thepage{} of \pageref*{LastPage}}
\makeatother

\begin{document} 

 \title{Planetary companions orbiting the M dwarfs GJ~724 and GJ~3988}
 \subtitle{A CARMENES and IRD collaboration \thanks{The RV data used in this work are only available in electronic form at the CDS via anonymous ftp to cdsarc.u-strasbg.fr (130.79.128.5)
or via \url{http://cdsweb.u-strasbg.fr/cgi-bin/qcat?J/A+A/} }}

    \author{P.~Gorrini\inst{\ref{iag}} 
      \and J.~Kemmer\inst{\ref{lsw}}
      \and S.~Dreizler \inst{\ref{iag}}
      \and R.~Burn \inst{\ref{mpia}}
      \and T.~Hirano \inst{\ref{japan-center}, \ref{japan-obs}, \ref{japan-uni}}
      \and F.\,J.~Pozuelos\inst{\ref{iaa}}
      \and M.~Kuzuhara \inst{\ref{japan-center}, \ref{japan-obs}}
      \and J.\,A.~Caballero \inst{\ref{cab}}
      \and P.\,J.~Amado \inst{\ref{iaa}}
      \and H.~Harakawa \inst{\ref{subaru}}
       \and T.~Kudo \inst{\ref{subaru}}
      \and A.~Quirrenbach \inst{\ref{lsw}}
      \and A.~Reiners \inst{\ref{iag}}
      \and I.~Ribas \inst{\ref{ice},\ref{ieec}}
      \and V.\,J.\,S.~B\'ejar \inst{\ref{iac}, \ref{la-laguna}}
      \and P.~Chaturvedi \inst{\ref{tls}}
      \and C.~Cifuentes \inst{\ref{cab}}
      \and D.~Galad\'i-Enr\'iquez \inst{\ref{caha}}
      \and A.\,P.~Hatzes \inst{\ref{tls}}
      \and A.~Kaminski \inst{\ref{lsw}}
      \and T.~Kotani \inst{\ref{japan-center}, \ref{japan-obs}, \ref{japan-uni}}
      \and M.~K\"urster \inst{\ref{mpia}}
      \and J.\,H.~Livingston \inst{\ref{japan-center}, \ref{japan-obs}, \ref{japan-uni}}
      \and M.\,J.~L\'opez~Gonz\'alez \inst{\ref{iaa}}
      \and D.~Montes \inst{\ref{ucm}}
      \and J.\,C.~Morales \inst{\ref{ice},\ref{ieec}}
      \and F.~Murgas \inst{\ref{iac}, \ref{la-laguna}}
      \and M.~Omiya \inst{\ref{japan-center}, \ref{japan-obs}}
      \and E.~Pall\'e \inst{\ref{iac}, \ref{la-laguna}}
      \and E.~Rodr\'iguez \inst{\ref{iaa}}
      \and B.~Sato \inst{\ref{japan-inst}}
      \and A.~Schweitzer \inst{\ref{hs}}
      \and Y.~Shan \inst{\ref{iag}, \ref{norway}}
      \and T.~Takarada \inst{\ref{japan-center}, \ref{japan-obs}}
      \and L.~Tal-Or \inst{\ref{ariel}, \ref{iag}}
      \and M.~Tamura \inst{\ref{uni-tokyo}, \ref{japan-center}, \ref{japan-obs}}
      \and S.~Vievard \inst{\ref{subaru}}
      \and M.\,R.~Zapatero~Osorio \inst{\ref{cab}}
      \and M.~Zechmeister \inst{\ref{iag}}
          }

    \institute{Institut f\"{u}r Astrophysik und Geophysik, Georg-August-Universit\"{a}t,  Friedrich-Hund-Platz 1, 37077 G\"{o}ttingen, Germany \label{iag}\\
    \email{paula.gorrini@uni-goettingen.de} 
    \and Landessternwarte, Zentrum f\"ur Astronomie der Universit\"{a}t Heidelberg, K\"{o}nigstuhl 12, 69117 Heidelberg, Germany \label{lsw}
    \and Max-Planck-Institut f\"ur Astronomie, K\"onigstuhl 17, 69117 Heidelberg, Germany \label{mpia}
    \and Astrobiology Center, 2-21-1 Osawa, Mitaka, Tokyo 181-8588, Japan \label{japan-center}
    \and National Astronomical Observatory of Japan, 2-21-1 Osawa, Mitaka, Tokyo 181-8588, Japan \label{japan-obs}
    \and Department of Astronomical Science, The Graduate University for Advanced Studies, SOKENDAI, 2-21-1 Osawa, Mitaka, Tokyo 181-8588, Japan \label{japan-uni}
    \and Instituto de Astrof\'isica de Andaluc\'ia (IAA-CSIC), Glorieta de la Astronom\'ia s/n, 18008 Granada, Spain \label{iaa}
    \and Centro de Astrobiolog\'ia (CSIC-INTA), ESAC Campus, Camino Bajo del Castillo s/n, 28692 Villanueva de la Ca\~nada, Madrid, Spain \label{cab}
    \and Subaru Telescope, 650 North A’ohoku Place, Hilo HI 96720, USA \label{subaru}
    \and Institut de Ci\`encies de l’Espai (ICE, CSIC), Campus UAB, Can Magrans s/n, 08193 Bellaterra, Spain \label{ice}
    \and Institut d’Estudis Espacials de Catalunya (IEEC), 08034 Barcelona, Spain \label{ieec}
    \and Instituto de Astrof\'isica de Canarias, V\'ia L\'actea s/n, 38205 La Laguna, Tenerife, Spain \label{iac}
    \and Departamento de Astrof\'isica, Universidad de La Laguna, 38206 La Laguna, Tenerife, Spain \label{la-laguna} 
    \and Th\"uringer Landessternwarte Tautenburg, Sternwarte 5, 07778 Tautenburg, Germany \label{tls}
    \and Centro Astron\'omico Hispano en Andaluc\'ia (CAHA), Observatorio de Calar Alto, Sierra de los Filabres, 04550 G\'ergal, Almer\'ia, Spain \label{caha}
    \and Departamento de F\'{i}sica de la Tierra y Astrof\'{i}sica and IPARCOS-UCM (Instituto de F\'{i}sica de Part\'{i}culas y del Cosmos de la UCM), Facultad de Ciencias F\'{i}sicas, Universidad Complutense de Madrid, 28040, Madrid, Spain \label{ucm}
    \and Department of Earth and Planetary Sciences, School of Science, Tokyo Institute of Technology, 2-12-1 Ookayama, Meguroku, Tokyo 152-8551, Japan \label{japan-inst}
    \and Hamburger Sternwarte, Universit\"at Hamburg, Gojenbergsweg 112, 21029 Hamburg, Germany \label{hs}
    \and Centre for Earth Evolution and Dynamics, Department of Geosciences, University of Oslo, Sem Saelands vei 2b 0315 Oslo, Norway \label{norway}
    \and Department of Physics, Ariel University, Ariel 40700, Israel \label{ariel}
    \and Department of Astronomy, Graduate School of Science, The University of Tokyo, 7-3-1 Hongo, Bunkyo-ku, Tokyo 113-0033, Japan \label{uni-tokyo}
    }

   \date{Received 6 June 2023 / Accepted 15 September 2023} 

 
  \abstract
   {We report the discovery of two exoplanets around the M dwarfs GJ~724 and GJ~3988 using the radial velocity (RV) method. We obtained a total of 153 3.5\,m Calar Alto/CARMENES spectra for both targets and measured their RVs and activity indicators. We also added archival ESO/HARPS data for GJ~724 and infrared RV measurements from Subaru/IRD for GJ~3988. We searched for periodic and stable signals to subsequently construct Keplerian models, considering different numbers of planets, and we selected the best models based on their Bayesian evidence. Gaussian process (GP) regression was included in some models to account for activity signals. For both systems, the best model corresponds to one single planet. The minimum masses are $10.75^{+0.96}_{-0.87}$ and $3.69^{+0.42}_{-0.41}$ Earth-masses for GJ~724\,b and GJ~3988\,b, respectively. Both planets have short periods ($P < \SI{10}{\day}$) and, therefore, they orbit their star closely ($a < \SI{0.05}{\astronomicalunit}$).  GJ~724\,b has an eccentric orbit ($e = 0.577^{+0.055}_{-0.052}$), whereas the orbit of GJ~3988\,b is circular. The high eccentricity of GJ~724\,b makes it the most eccentric single exoplanet (to this date) around an M dwarf. Thus, we suggest a further analysis to understand its configuration in the context of planetary formation and architecture. In contrast, GJ~3988\,b is an example of a common type of planet around mid-M dwarfs.}

   \keywords{planetary systems -- techniques: radial velocities -- stars: individual: GJ~724, GJ~3988 -- stars: late-type -- planets and satellites: detection  
               }

   \maketitle
%

\section{Introduction}
\label{sec:intro}

The number of detected exoplanets is growing thanks to the advances in astronomical instrumentation. High-precision spectrographs are capable of detecting radial velocity (RV) signals on the order of \SI{1}{\meter\per\second} (e.g. HARPS, \citealt{Mayor2003}; CARMENES, \citealt{Quirrenbach2014}) and even lower (e.g. ESPRESSO, \citealt{Pepe2021}; MAROON-X,\citealt{Seifahrt2018}). In this sense, the detection of terrestrial exoplanets around low-mass stars, such as M dwarfs, is relatively easily  attainable and, therefore, advantageous. This is due to the fact that these types of stars have lower masses (0.08--0.60\,M$_{\odot}$) and are smaller (0.1--0.6\,R$_{\odot}$) when compared to Sun-like stars, thus producing larger RV amplitude variations for a given planetary mass.

However, planetary signals can be mimicked by stellar activity (e.g. \citealt{Robertson2015}; \citealt{Lubin2021}; \citealt{Gorrini2022}), especially in 
M dwarfs, as they tend to be magnetically active (e.g. \citealt{Reiners2012, Reiners2022}; \citealt{Mignon2023}). Therefore, it is necessary to treat this phenomenon when searching for exoplanets. As stellar activity can manifest itself in the stellar rotation period and its harmonics \citep{Boisse2011}, it is necessary to have a clear determination of the stellar rotation period and analyse  the signals that appear on its integer fractions in depth.  One successful method for treating stellar activity is to model it with Gaussian process (GP) regression (e.g. \citealt{Haywood2014}; \citealt{Rajpaul2015}; \citealt{Angus2018}; \citealt{Stock2020a}; \citealt{Kossakowski2022}; \citealt{Kemmer2022}), but this must be used with caution, that is, by ensuring a proper (physical) training to avoid biased results, such as over-fitting or absorbing planetary signals (see e.g. \citealt{Cabot2021}; \citealt{Rajpaul2021}; \citealt{Stock2023}). 

Another aspect that has to be taken into consideration when looking for planetary signals is a possible mean motion resonance (MMR) for systems with highly eccentric orbits, such as a 2:1 MMR can mimic an RV signal of a single high-eccentricity planet (e.g. \citealt{Anglada-Escude2010}; \citealt{Wittenmyer2013}; \citealt{Kurster2015}; \citealt{Nagel2019}), leading to erroneous conclusions regarding the number of planets in the system. This is important when characterising planetary systems, as the planet's orbital solution can change drastically, coupled with our current knowledge of the incidence of eccentric and circular planetary orbits. 



\begin{table*}
    \caption{Stellar properties and parameters of the two planet-hosting stars.}
    \label{tab:stellar-properties}
    \centering
    \begin{tabular}{l c c  r}
        \hline\hline
        \noalign{\smallskip}
        Parameter                 & GJ~724               & GJ~3988                          & Reference      \\
        \noalign{\smallskip}
        \hline
        \noalign{\smallskip}
        Name                        & \object{BD--13 5069} & \object{G 203--42} & \citet{Schonfeld1886, Giclas1971}\\
        Karmn                     & J18409--133           & J17033+514                       & \citet{Caballero2016b}    \\
        Spectral type             & M1.0\,V                & M4.5\,V                            & \citet{Reid1995}    \\
        $\alpha$ (J2000)                & 18:40:57.31          & 17:03:23.88                      & \citet{GaiaCollaboration2022}      \\ 
        $\delta$ (J2000)               & $-$13:22:46.6        & $+$51:24:22.9                    & \citet{GaiaCollaboration2022}      \\
        $G^{(a)}$ (\si{\mag})             & $9.7722 \pm 0.0028  $ & $ 11.9810 \pm 0.0028$               & \citet{GaiaCollaboration2021}          \\
        $J^{(a)}$ (\si{\mag})             & $7.397 \pm 0.018 $  &  $8.768 \pm 0.027$                & \citet{Skrutskie2006}\\
        $\pi$ (\si{mas})            & $58.841 \pm 0.030$     & $100.927 \pm 0.022$            & \citet{GaiaCollaboration2021}      \\
        $d$ (\si{pc})             & $16.9949 \pm 0.0086$   & $9.9081 \pm 0.0022$                & \citet{GaiaCollaboration2021}      \\
        $\mu_\alpha \cos{\delta}$ (\si{\mas\,\per\year})  & $-94.287 \pm 0.032$ & $+124.410 \pm 0.025$  &  \citet{GaiaCollaboration2021} \\
        $\mu_\delta$ (\si{\mas\,\per\year})  &  $-671.266 \pm 0.027$ & $+610.145 \pm 0.028$      & \citet{GaiaCollaboration2021}           \\
        $\gamma$ (km\,s$^{-1}$) & $-33.36 \pm 0.22$ & $+37.17 \pm 0.30$ & \citet{GaiaCollaboration2021}      \\
        $U$ (km\,s$^{-1}$) & $-14.5221 \pm 0.0.0076$ & $-21.0460 \pm 0.0068$ & This work \\
        $V$ (km\,s$^{-1}$) & $-60.124 \pm 0.023$ & $+38.956 \pm 0.014$ & This work \\
        $W$ (km\,s$^{-1}$) & $-15.754 \pm 0.013$ & $17.257 \pm 0.011$ & This work \\
        $T_{\text{eff}}^{(b)}$ (\si{K}) & $3799 \pm 81$        & $3273 \pm 101$                    & \citet{Marfil2021}   \\
        $\log g$                     & $4.88 \pm 0.08$      & $4.92 \pm 0.14$                  & \citet{Marfil2021}    \\
         \text{[Fe/H]} (dex)                    & $-0.02 \pm 0.05$     & $-0.12 \pm 0.17$                 & \citet{Marfil2021}    \\
        $L_\star$ (L$_{\odot}$)  & $0.05118 \pm 0.00034$ & $0.004026 \pm 0.000020$       & This work     \\
        $R_\star$ (R$_{\odot}$)      & $0.52226 \pm $ 0.02234    & $0.19734 \pm 0.01219$                & This work     \\
        $M_\star$ (M$_{\odot}$)        & $0.5271 \pm 0.0263$      &    $0.1842 \pm 0.0153$             & This work     \\
        $v\sin{i}$ (\si{\km/\second})                  &    $ < 2$     &    $< 2$     &     \citet{Reiners2018}                  \\
        pEW(H$\alpha$) (\AA) & $+0.454 \pm 0.010$ &  $+0.071 \pm 0.042$ & \citet{Fuhrmeister2020}\\
        $<B>$ (G) & $310 \pm 60$ & $< 520$ & \citet{Reiners2022} \\
        $\log{R'_{\rm HK}}$ (dex) & $-4.641^{+0.055}_{-0.062}$ & ... & \citet{Perdelwitz2021} \\
        $<P_\text{rot}>$ (\si{\day})        & $57 \pm 1$ / $\sim 28$                     & $116\pm3$             & This work \\
        Gal. population & Thin disc & Thin disc & This work \\
        \noalign{\smallskip}
        \hline
    \end{tabular}
    \tablefoot{
    \tablefoottext{a}{See \citet{Cifuentes2020} for a photometry compilation from Johnson $B$ to {\em WISE} $W4$.}
    \tablefoottext{b}{$T_{\rm eff}$ uncertainty is the sum of the measurement error given in \citet{Marfil2021} and the weighted average of the widths of the differences to the literature values given in Table 5 of \citet{Marfil2021} serving as an estimate for the systematic error.}
    }   
\end{table*}

In this paper, we report planets orbiting two stars, GJ~724 (M1.0\,V) and GJ~3988 (M4.5\,V), observed with CARMENES as a part of the guaranteed time observations to search for exoplanets around M dwarfs \citep{Ribas2023}. 
GJ~3988 was also observed with the InfraRed Doppler (IRD) instrument at the Subaru telescope. 
In \autoref{sec:properties}, we summarise the stellar properties of the host stars, while in \autoref{sec:data}, we describe our data set and measurements. We explain the methods used in the analysis in \autoref{sec:methods}.  In \autoref{sec:rot}, we determine the mean stellar rotation period for each star. We analyse all datasets in \autoref{sec:analysis} and discuss the implications of our study in \autoref{sec:discussion}. Finally,  we summarise our conclusions in \autoref{sec:conclusions}.

\section{Stellar properties}
\label{sec:properties}

In spite of the difference in spectral types, which translate into distinct stellar effective temperatures ($T_{\rm eff}$), radii ($R_\star$), and masses ($M_\star$), the two planet-hosting stars have many parameters in common.
While GJ\,724 was already tabulated in the Bonner Durchmusterung \citep{Schonfeld1886}, this star and GJ\,3988 were (re-)discovered in high proper-motion surveys by \citet{Ross1928} and \citet{Giclas1971}, respectively. The farther distance of the former star partially compensates its earlier spectral type and, therefore, it is not very much brighter than the latter.
However, their distances are so short (from \SI{9.9}{\pc} to \SI{17}{\pc}) that their classification as solar neighbours is assumed, supported by the wealth of public data available.

In \autoref{tab:stellar-properties}, we provide a summary of the most important stellar properties for each star. First, we retrieved the spectral types, coordinates, magnitudes ($G$ and $J$), parallactic distances, proper motions, and radial and rotation velocities from the literature for both GJ~724 and GJ~3988.  There are numerous independent spectral type determinations that are consistent with the tabulated ones, between M0 and M1 for GJ~724 \citep[e.g.][]{Bidelman1985, Gaidos2014} and between M4 and M5 for GJ~3988 \citep[e.g.][]{Lepine2013, Terrien2015}. Something similar happens to the stellar atmospheric parameters \citep[e.g.][]{Passegger2018,Rajpurohit2018, Kuznetsov2019, Khata2021, Ishikawa2022}.
In our case, we used the values of $T_{\rm eff}$, $\log{g}$, and [Fe/H] from CARMENES VIS and NIR template spectra determined by \citealt{Marfil2021} with the {\tt SteParSyn} code\footnote{\url{ https://github.com/hmtabernero/SteParSyn/}} \citep{Tabernero2022}.
The bolometric luminosities ($L_\star$) were measured from the integration of the spectral energy distribution following \cite{Cifuentes2020} but with updated {\em Gaia} DR3 photometric and astrometric data \citep{GaiaCollaboration2022}.  The radii and masses followed from $L_\star$, Stefan-Boltzmann's law, and the $M_\star$-$R_\star$ relation by \cite{Schweitzer2019}.  We also computed galactocentric space velocities from the available data, which put both GJ~724 and GJ~3988 in the thin disc galactic population and measured mean rotation periods by fitting GPs to the photometric data as described in \autoref{sec:rot}. 
The kinematics and long rotation periods are consistent with the pseudoequivalent width of the H$\alpha$ line in faint absorption \citep{Fuhrmeister2020}, stringent upper limits to the rotational velocities imposed by the CARMENES spectral resolution \citep{Reiners2018}, no detection of X-ray flux \citep{Voges1999, Stelzer2013}, and (only for GJ~724) faintly measured magnetic fields and Ca~{\sc ii} emission \citep{Astudillo-Defru2017a, Perdelwitz2021, Reiners2022}. The ages of both stars can be poorly constrained from these data, but are very likely of the order of \SI{4}{\giga\year} \citep[][although gyrochonology at these late spectral types is not reliable; e.g \citealt{DiezAlonso2019}]{Dungee2022}. Further details on stellar properties determination were provided by \citet{Caballero2022}.

Finally, we address the potential multiplicity of GJ~724 and GJ~3988.
Both stars were observed with the lucky imager FastCam at the 1.5\,m Telescopio Carlos S\'anchez \citep{Jodar2013, Cortes-Contreras2017}.
Besides, since GJ\,3988 is located at $d <$ \SI{10}{\pc}, it has also been the subject of numerous high-resolution imaging surveys for companions at close and very close separations (down to \SI{\sim 50}{\mas}) with other lucky imagers \citep[AstraLux at 2.2\,m Calar Alto;][]{Janson2014}, coronographs in the near infrared \citep[2.3\,m Bok telescope at Steward Observatory;][]{McCarthy2004}, adaptive optics systems \citep[Robo-AO;][]{Lamman2020}, and even the {\em Hubble Space Telescope} \citep[with NICMOS and the F180M band;][]{Dieterich2012}. None of these surveys identified a close or very close companion candidate to GJ~724 and GJ~3988. Recently, \citet{Cifuentes2023} looked for additional companions with {\em Gaia} DR3 data at even closer separations, with the {\tt RUWE} parameter and other close-binarity astrophotometric indicators, and at wide and very wide separations of up to $2 \times 10^5$\,au with a common proper-motion and parallax survey.
Again, no companions were reported, so both GJ~724 and GJ~3988 appear to be single M dwarfs.


\section{Data}
\label{sec:data}

\begin{table*}
    \caption{Overview of the RV data.}
    \label{tab:rv-data}
    \centering
    \begin{tabular}{lccccccc}
        \hline \hline
        \noalign{\smallskip}
         Instrument & \multicolumn{2}{c}{Date}  & Time span                & Mean error  &   RMS & \multicolumn{1}{c}{$N_\textnormal{data}$}   \\
                            &  Begin      & End         &\multicolumn{1}{c}{[d]}   & [\si{\meter\per\second}]  & [\si{\meter\per\second}]      &                            \\
        \noalign{\smallskip}
        \hline
        \noalign{\smallskip}
            \multicolumn{5}{c}{\textit{GJ~724}}  \\
        \noalign{\smallskip} 
                 CARMENES   &  May 2016   & August 2022 & 2\,277                   & 1.94  &   5.48     & 83                                          \\
                 HARPS      &  April 2008 & March 2012  & 1\,424                   & 1.50  &   5.61     &  27                                         \\
        \hline    
        \noalign{\smallskip}
                    \multicolumn{5}{c}{\textit{GJ~3988}}  \\
        \noalign{\smallskip} 
                 CARMENES   &  April 2016 & August 2022 & 2\,313                   & 1.81  &  4.92      & 70                                          \\
                 IRD   &  February 2019 & July 2021 & 871                   & 2.89  &   8.19     & 94        \\
        \hline
    \end{tabular}
    
\end{table*}

\begin{table*}
    \caption{Overview of the photometric data. }
    \label{tab:phot-data}
    \centering
    \begin{tabular}{p{2.5cm}cccSSSS}
        \hline \hline
        \noalign{\smallskip}
        Instrument                   & \multicolumn{2}{c}{Date} & Band           & {Time span}             & {Mean error}              & {RMS}   & \multicolumn{1}{c}{$N_\textnormal{binned}$}       \\
                                     & Begin                    & End            &                       & \multicolumn{1}{c}{[d]} & {[ppt]} & {[ppt]}                                       &     \\
        \hline
        \noalign{\smallskip}
        \multicolumn{8}{c}{\textit{GJ~724}} \\
        \noalign{\smallskip}
        OSN                          & Aug. 2020              & Aug. 2021    & $V$                   & 364                     & 3.50 & 4.78                                       & 57  \\
                                  &               &     & $R$                   & 364                     & 3.71 & 4.41                                       & 57  \\
        TJO                          & Aug. 2020              & July 2021    & $R$                   & 314                     & 1.21 & 6.40                                       & 78  \\
        \hline
        \noalign{\smallskip}
        \multicolumn{8}{c}{\textit{GJ~3988}}  \\
        \noalign{\smallskip}                                                      
        TJO                          & Sept. 2022           & March 2023     & $R$                   & 178                     & 0.84  & 5.35                                        & 260 \\
        MEarth-tel07 (v6,v7,v9 to v12) & April 2012         & Oct. 2018     & RG715                  & 2746                    & 3.63 & 7.43                                       & 312 \\
        MEarth-tel08 (v5,v8,v10)       & Dec. 2012          & Sept. 2019    & RG715                  & 2465                    & 3.67 & 6.43                                       & 308 \\
        SuperWASP                    & April 2006           & Aug. 2008     &  (4000--7000\,\AA) & 853 & 10.95 & 512.17        & 260 \\
        \hline
    \end{tabular}
\end{table*}

\subsection{Spectroscopic data}
\label{subsec:rv-data}
The spectral data were obtained with CARMENES\footnote{Calar Alto high-Resolution search for M-dwarfs with Exoearths with Near-infrared and optical Échelle Spectrograph.} \citep[][]{Quirrenbach2014} for both GJ~724 and GJ~3988. We also incorporated RV data from the High Accuracy Radial velocity Planet Searcher spectrograph \citep[HARPS;][]{Mayor2003} for GJ~724. For GJ~3988, we used data from the InfraRed Doppler \citep[IRD;][]{Tamura2012, Kotani2018} at the Subaru Telescope. The main properties of both RV data sets are summarised in \autoref{tab:rv-data}. 


\subsubsection{CARMENES}
\label{subsubsec:carmenes}

GJ~724 and GJ~3988 were observed with CARMENES, installed at the 3.5\,m telescope at the Calar Alto observatory. It consists of two high-resolution spectrographs, covering the optical 5200--9600\,{\AA} (VIS) and near-infrared 9600--17100\,{\AA} (NIR) ranges,  with spectral resolutions of $\mathcal{R}=\num{94 600}$ and $\mathcal{R}=\num{80 400}$, respectively. For GJ~724, we made use of 83 VIS observations over a time span of \SI{2277}{\day} (\SI{\sim6}{\year}), from May 2016 to August 2022. As for GJ~3988, we used 70 VIS observations with a time span of \SI{2313}{\day}, from April 2016 to August 2022. 

We did not include the NIR CARMENES RVs in our analysis as the errors and RMS were much higher (for GJ~724 by a factor of 4 and 2, respectively; whereas for GJ~3988 this was a factor of 4 and 3, respectively) than the data in the VIS channel given the same time sampling. Therefore, adding this data set would not contribute in any way to our models and would, in fact, decrease the quality of the fits. We refer to \cite{Bauer2020} for a comparative of VIS and NIR CARMENES RV data of a nearby M dwarf.



All raw spectra were processed with the CARMENES pipeline \texttt{caracal} \citep{Caballero2016a}. The RVs were computed using the \texttt{serval} code \citep{Zechmeister2018}, which uses a high signal-to-noise (S/N) stellar template to perform a least-square fit to each observation, including corrections of the instrumental drift from simultaneous measurements of the Fabry-P\'erot interferometer and the Earth's barycentric velocity. As described by \citet{Tal-Or2019} and, especially, \citet[][see their Sect. 4.4]{Ribas2023}, we also corrected RVs for nightly zero points.

We also computed the stellar activity indicators defined by \citet{Zechmeister2018}, namely the chromatic index (CRX), which measures the wavelength dependence of the RV, and the differential line width (dLW), besides the H$\alpha$ index, Ca~{\sc ii} infrared triplet indices (Ca\,IRT\,a~$\lambda$8498\,{\AA}, Ca\,IRT\,b~$\lambda$8542\,{\AA}, and Ca\,IRT\,c~$\lambda$8662\,{\AA}), and sodium doublets (Na~D$_1$ and Na~D$_2$), also according to \citet{Zechmeister2018}. Moreover, using the methodology described by \citet{Schofer2019}\footnote{The wavelengths specified by \cite{Schofer2019} are in vacuum, while we state the values in air.}, we computed indices for calcium hydride \citep[CaH3;][]{Reid1995}, titanium oxide ($\text{TiO}\,7048\,\r{A}$, $\text{TiO}\,8428\,\r{A}$, $\text{TiO}\,8858\,\r{A})$, $\text{Fe}\,8689\,\r{A}$, $\text{VO}\,7434\,\r{A}$, $\text{He}\,10830\,\r{A}$, $\text{Pa}\beta$, and the FeH Wing-Ford band (WFB). Lastly, we computed cross-correlation (CCF) parameters as described by \citet{Lafarga2020}, such as the contrast (CCF-contrast) and the full width at half maximum (CCF-FWHM).

\subsubsection{HARPS}
\label{subsubsec:harps}
We also incorporated  the data from observations of GJ~724 with HARPS, installed at the 3.6\,m telescope located at La Silla observatory. It covers wavelengths between \SI{3780}{\angstrom} and \SI{6910}{\angstrom} and has a spectral resolution of \num{115000}. We made use of the RV data refined by \citet{Trifonov2020}, which were re-processed using  \texttt{serval} \citep{Zechmeister2018}, obtaining a better precision of the RV measurements when compared to the original {\tt ESO-DRS} pipeline. The data were directly retrieved from their online catalog \citep{Trifonov2020}. The number of RV data points is 27,  covering a time span of \SI{1424}{\day} (\SI{\sim4}{\year}), from April 2008 to March 2012, and, therefore, with no overlap with the CARMENES data.

\subsubsection{IRD}
We observed GJ~3988 with the IRD instrument atop the Subaru 8.2\,m telescope as part of the Subaru Strategic Program (SSP), which searches for exoplanets around mid-to-late M dwarfs by a blind Doppler survey. A total of $98$ near-IR spectra (9500--17300\,\AA) were secured between February 2019 and July 2021. The integration time for each spectrum varied from \SI{220}{\second} to \SI{1800}{\second}, depending on the weather condition. We also injected light from the laser-frequency comb (LFC) into a reference fiber, which was used to trace the instantaneous instrumental profile of the spectrograph. 

Wavelength-calibrated one-dimensional spectra for both stellar and LFC fibers were extracted with the standard IRD reduction pipeline \citep{Kuzuhara2018}. The typical S/N for the stellar spectra was 65--100\,\si{\per\pixel} at \SI{10000}{\AA}. We measured RVs by fitting the individual spectra to a template spectrum of GJ~3988, which was also extracted from the observed IRD data, with the forward modeling technique \citep{Hirano2020}. The typical RV internal error was 2--3\,\si{\meter\per\second} for each data point. 

There are systematic offsets in the IRD RV measurements that are specific to each observing sequence (IRD observations are generally conducted on bright nights, so there are gaps between each sequence of bright nights). The offsets may be due to brightness variations of the LFC or other instrumental instabilities. We plotted the observing epochs versus the RV measurements of GJ~3988 in order to examine the offsets, to later correct them with the following post-processing. We grouped RV data points in each epoch with multiple targets being combined and computed the median of RV measurements in each group, allowing us to cancel signals from potential planets and stellar activity, leaving only instrumental signals. Finally, to apply the correction, we subtracted these median RV values from the individual RVs in each group. For this target, the central 80th percentile range of the offset values is $-2$ \si{\meter\per\second} to $+2$ \si{\meter\per\second}.

The observing strategy from IRD is to typically take multiple (short) exposures on a single night, with relatively low S/N values. Since we are not interested in short periods, but in getting observations comparable with CARMENES, we binned the data nightly. The binning resulted in a total of 50 IRD RV data points with a mean error and RMS of \SI{2.81}{\meter\per\second} and \SI{5.99}{\meter\per\second}, respectively.

\subsection{Photometric data}

We used photometric data from multiple sources for our analysis. An overview of the available data for each star can be found in \autoref{tab:phot-data}. Outliers were removed by applying a $3\sigma$ clipping to all photometric datasets, which were afterwards binned daily. 
In the following, we describe the individual instruments and how the data were taken.

\label{subsec:phot-data}
\subsubsection{OSN}
\label{subsubsec:phot-OSN}
Photometric CCD observations for GJ~724 were collected at the Observatorio de Sierra Nevada (OSN) in Granada, Spain, with the T90 telescope, which is a 90\,cm Ritchey-Chr'etien telescope equipped with a CCD camera VersArray 2k$\times$2k with a field of view of 13.2$\times$13.2\,arcmin$^2$ 
The camera is based on a high quantum efficiency back-illuminated CCD chip, type Marconi-EEV CCD42-4, with optimised response in the ultraviolet \citep{Amado2021}. The observations were collected in both Johnson $V$ and $R$ filters on 57 nights, during two runs: run 1 consisted of 24 epochs obtained during the period August-October 2020, while run 2 consisted of 33 epochs obtained during the period June-August 2021. 
Each epoch typically consisted of 20 exposures in each filter per night, of \SI{30}{\second} and \SI{20}{\second} in the $V$ and $R$ filters, respectively. 

All CCD measurements were obtained by the method of synthetic aperture photometry using no binning. Each CCD frame was corrected in a standard way for bias and flat fielding. Different aperture sizes were also tested in order to choose the best one for our observations. A number of nearby and relatively bright stars within the frames were selected as check stars in order to choose the best ones to be used as reference stars. 

\subsubsection{TJO}
\label{subsubsec:phot-TJO}
We include photometric observations of GJ~724 and GJ~3988 from the 0.8\,m Telescopi Joan Or\'o (TJO) at the Parc Astr\`onomic del Montsec, Lleida, Spain. We made use of the Johnson $R$ filter of the LAIA imager, a 4k$\times$4k CCD with a field of view of 30$\times$30\,arcmin$^2$ 
and a scale of 0.4\,arcsec\,pixel$^{-1}$. 
The images were calibrated with darks, bias and flat fields with the {\tt ICAT} pipeline \citep{Colome2006} of the TJO. The differential photometry was extracted with {\tt AstroImageJ} \citep{Collins2017} using the aperture size that minimised the root mean square (RMS) of the resulting relative fluxes, and a selection of the 30 brightest comparison stars in the field that did not show variability. Then, we used our own pipelines to remove outliers and measurements affected by poor observing conditions or presenting a low S/N.

\subsubsection{MEarth}
\label{subsubsec:phot-Mearth}
For GJ~3988, we included publicly available data from the MEarth Project \citep[MEarth;][]{Berta2012, Irwin2015}, corresponding to the {Data Release 11} (DR11). MEarth has two different facilities: MEarth-North, located at the Fred Lawrence Whipple Observatory in Arizona, USA, and MEarth-South, located at Cerro Tololo Inter-American Obervatory in Chile. We made use of data from MEarth-North, which consists of eight near-identical telescope systems, with 0.4\,m $f$/9 Ritchey-Chr\'etien Cassegrain design on a German equatorial mount. The pixel scale is approximately 0.76\,arcsec\,pixel$^{-1}$. 
The filter is fixed and consists of $2\times$\SI{1.5}{\milli\meter} plates of RG715 Schott glass built into the camera housing itself.

Because of possible offsets and other unknown effects, we divided the MEarth light curves into the individual instrument sections denoted in the light curve files. Specifically, we used telescope 7 (version 6, 7, and 9 to 12), and telescope 8 (versions 5, 8, 10), as the other data sets contained only a few data points ($<15$) or had short baselines (\SI{< 50}{\day}) and, therefore, the inclusion of these few points may result in higher uncertainties due to unknown offsets between the instrument versions. 


\subsubsection{SuperWASP}
\label{subsubsec:phot-SWASP}
GJ~3988 was also monitored by the Super-Wide Angle Search for Planets \citep[SuperWASP;][]{Pollacco2006}. The survey comprises two identical robotic telescopes, at La Palma, Spain, and Sutherland, South Africa, each with eight cameras observing through broadband filters (4000--7000\,\AA). We used data spanning April 2006 to August 2008 from the first public data release \citep{Butters2010}, totalling 260 daily-binned epochs over three seasons. The light curves were extracted via aperture photometry by the SuperWASP team, and corrected for instrumental systematics using the method of \citealt{Tamuz2005}. 

\section{Methods}
\label{sec:methods}

\subsection{Periodogram analysis}
To search for periodic signals in the available RVs, activity indicators, and photometry, we used the generalised Lomb-Scargle periodogram \citep[{\tt GLS};][]{Zechmeister2009} implemented in  \texttt{exostriker} \citep{Trifonov2019} and \texttt{astropy} \citep{AstropyCollaboration2013,AstropyCollaboration2018,AstropyCollaboration2022}. The periodograms were normalised by the residuals of the data following \cite{Zechmeister2009} and the false-alarm probabilities (FAPs) were calculated using the analytic expression from \cite{Baluev2008}. Our plots show the FAP levels of \SI{10}{\percent}, \SI{1}{\percent}, and \SI{0.1}{\percent}; we consider a signal to be significant if it has a FAP lower than \SI{1}{\percent}. The temporal stability of signals was assessed using the stacked Bayesian {\tt GLS} peridogram \citep[{\tt s-BGLS};][]{Mortier2015, Mortier2017}.

\begin{figure*}
    \centering
    \includegraphics[width=\columnwidth]{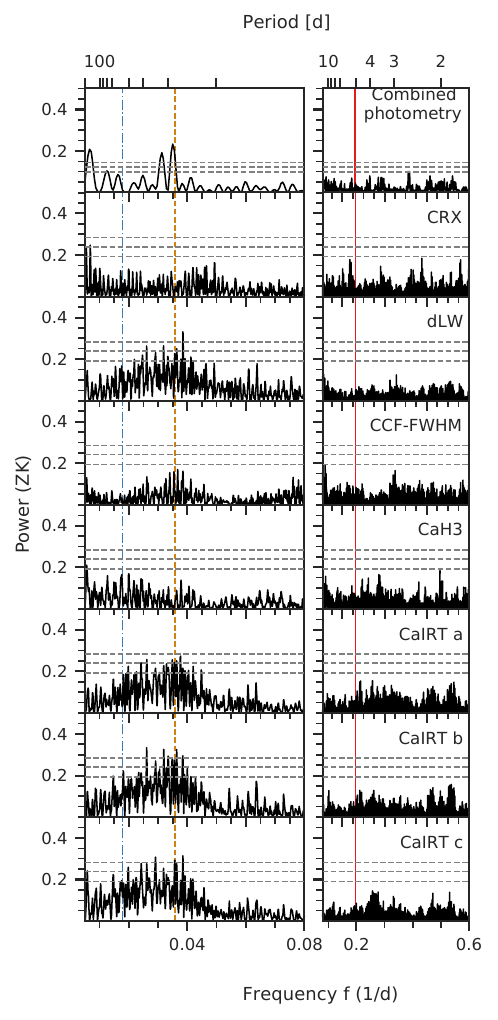}
    \includegraphics[width=\columnwidth]{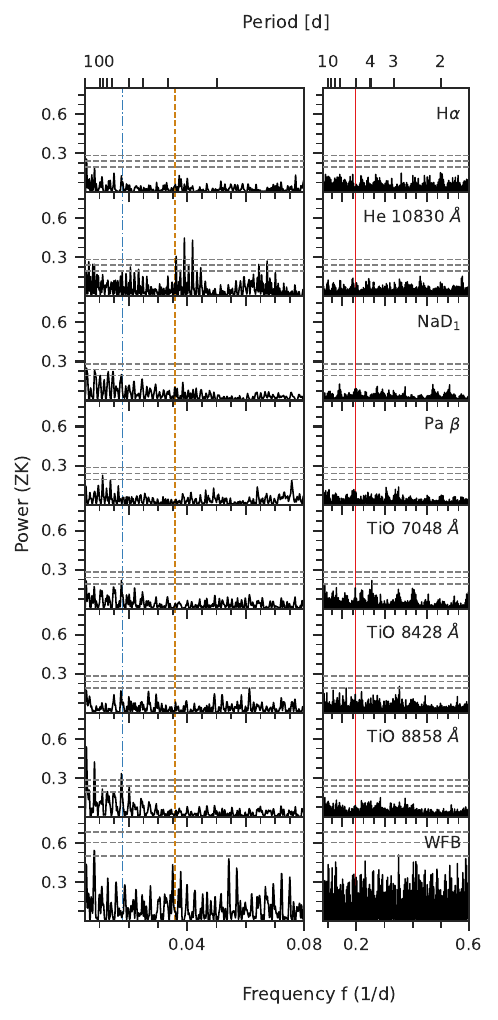} 
    \caption{GLS periodograms of the (combined) photometric data and spectroscopic activity indicators from GJ~724 with peaks that reach at least a FAP level of 10\% in the period range of \SI{2}{\day} to \SI{200}{\day}. The gray dashed horizontal lines correspond to the FAP levels of 0.1\%, 1\%, and 10\% (from top to bottom, respectively). The mean rotation period of \SI{56}{\day} determined from the photometry is marked by the blue dashed line, while its second harmonic (${P_{rot}}/2$) of \SI{28}{\day} is depicted by the orange dashed line. The period of the 5.1-day planet is highlighted by the red solid line.}
    \label{fig:activity_GLS_GJ724}
\end{figure*}

\subsection{Modelling}
\label{subsec:modelling}
For all modelling tasks, we used the \texttt{juliet} package \citep{Espinoza2019}. It combines the publicly available packages \texttt{radvel} \citep{Fulton2018} for fitting RVs and \texttt{batman} \citep{Kreidberg2015} for fits of photometry. For the sampling within \texttt{juliet}, we chose the dynamic nested sampling algorithm \citep{Higson2019} from \texttt{dynesty} \citep{Speagle2020, Koposov2022}, which allowed us to determine the Bayesian evidence for our models. We assumed that a model performs significantly better if the difference in $\ln\mathcal{Z}$ compared to the competing model was greater than \num{5}. Models with $\Delta\ln\mathcal{Z}>2.5$ were considered to be moderately preferred, and all models below that threshold were considered indistinguishable \citep[following e.g.][]{Trotta2008}. In addition to the static Keplerian model components, \texttt{juliet} provides GP as a red noise component using the \texttt{celerite} \citep{Foreman-Mackey2017} and \texttt{george} \citep{Ambikasaran2015} packages. As GP models can flexibly account for correlated noise, we compared the models with a GP separately from those without a GP (only Keplerian).

We used three different kernels to model stellar activity present in our data on the basis of a case-by-case examination: the double simple harmonic oscillator kernel \citep[dSHO;][]{David2019, Gillen2020}, as described by \cite{Kossakowski2021}; the quasi-periodic kernel \citep[QP;][]{Haywood2014, Rajpaul2015}, following the parametrisation outlined by \cite{Espinoza2019}, and lastly the quasi-periodic cosine kernel \citep[QPC;][]{Perger2021}, which is the sum of the QP kernel and a cosine function. The latter is parameterised by the same hyper-parameters as the QP kernel but also including the amplitudes of the QP component and the cosine function.
An overview of the parameters and standard priors used for the GP modelling is given in \autoref{tab:priors_GPs}.


\section{Rotation periods}
\label{sec:rot}




\subsection{GJ~724}
\label{subsec:Prot-GJ724}
\subsubsection{Spectroscopic rotation period}
\label{subsubsec:Prot-GJ724-spec}

In \autoref{fig:activity_GLS_GJ724} we show the periodograms of the photometric data and the activity indicators of GJ~724 that reach at least a $\text{FAP}$ level below \SI{10}\percent{}. The most significant signal (FAP $<$\SI{0.1}{\percent}) from the activity periodograms corresponds to \SI{\sim180}{\day} in the TiO $8858$\AA\, indicator, which is also present in the CRX, CaH$3$, Ca\,IRT\,c, $\text{TiO}\,7048\,\r{A}$, $\text{Na~D}_1$, and WFB indices (albeit not at  a significant level). This signal is the second harmonic of the period of one year and is therefore likely to be caused by the contamination of the spectra. 

Other signals with a FAP below \SI{0.1}{\percent} are found around \SI{122}{\day} and \SI{56}{\day} in the $\text{TiO}\,8858 \r{A}$ index. The former signal is also a harmonic of a year ($P_{1/3}$), whereas the \SI{56}{\day} period appears as well in other indicies ($\text{Na~D}_1$, Ca\,IRT\,b and Ca\,IRT\,c, $\text{TiO}\,7050$, $\text{He}\,\lambda10830\,\r{A}$, and WFB) with a FAP level between \SI{10}{\percent} and \SI{1}{\percent}. Half of this periodicity, that is \SI{28}{\day}, is also present in other activity indicators (dLW, Ca\,IRT\,a, Ca\,IRT\,b, Ca\,IRT\,c, $\text{H}_\alpha$, and $\text{He}\,\lambda10830\,\r{A}$), being only significant in the Ca\,IRT\,b and Ca\,IRT\,c indices. While this signal is close to the moon cycle, it matches also the second harmonic of the \SI{56}{\day} signal. 

There are also many other significant peaks between \SI{20}{\day} and \SI{40}{\day} (dLW, Ca\,IRT\,b, Ca\,IRT\,c, and $\text{He}\,10830\r{A}$). In order to check if they originate from a common period related to stellar activity, we performed a GP fit to the activity indicators (separately) using the dSHO kernel with uniform priors on the period from \SI{2}{\day} to \SI{100}{\day}. We only get a clear detection of the Ca\,IRT\,c index, with a period around \SI{\sim 50}{\day}. The Ca\,IRT\,a and Ca\,IRT\,b indices resulted in bimodal distributions of the period with values of \SI{\sim28}{\day} and \SI{\sim56}{\day}. By constraining the period of the TiO $\lambda 8858$\AA\, index to \SI{40}{\day} to \SI{60}{\day}, where the most significant signals are for this indicator (excluding the one year harmonic), we obtained a period of around \SI{57}{\day}. In this sense, we cannot retrieve an accurate rotation period from the activity indicators.

\subsubsection{Photometric rotation period}
\label{subsubsec:Prot-GJ724-phot}

\begin{figure}
    \centering
    \includegraphics[width=\columnwidth]{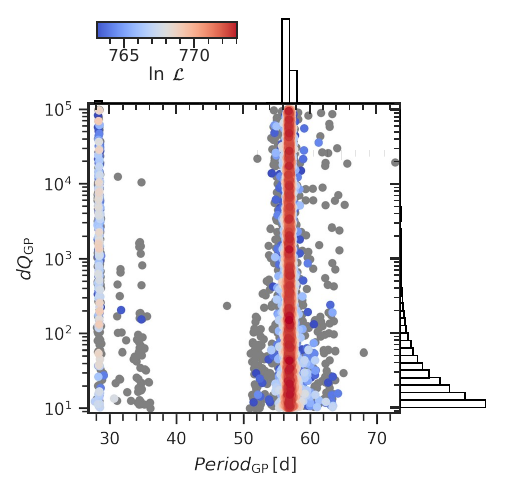}
    \includegraphics[width=\columnwidth]{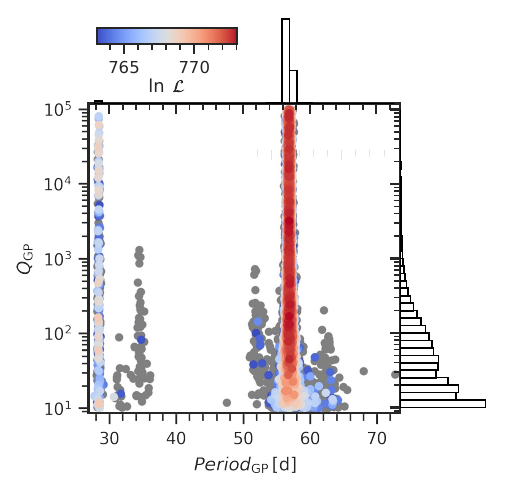}
    \caption{Posterior distribution from $dQ_\text{GP}$ against $P_{GP}$ (top) and $Q_\text{GP}$ against $P_{GP}$ (bottom) from the GP fit to the photometric data of GJ~724 using a dSHO kernel. The colour bar displays the log-likelihood, while gray points correspond to the samples with $\Delta \ln \mathcal{L} > 10$.}
    \label{fig:dQ-Prot_GJ724}
\end{figure}

The GLS periodogram of the (combined) OSN and TJO photometric data is also displayed in \autoref{fig:activity_GLS_GJ724}. The strongest peak is at \SI{\sim 28}{\day} with a  $\text{FAP}$ of $ \SI{2.9e-6}{\percent}$, followed by two slightly weaker signals ($\text{FAP} > \SI{1e-5}{\percent} $) at around \SI{147}{\day} and \SI{32}{\day}.
We performed a GP on the photometric data in order to retrieve the stellar rotation period of GJ~724 using the dSHO kernel.  All the GP parameters were shared between the different datasets, except for the amplitude. We searched for periodicities between 2 and 200 days, using a uniform prior on the periodic component. There is a detection on the posterior distribution of the GP rotation period parameter of \SI{28.4\pm0.1}{\day}, but for the parameters $Q0$ and $dQ$ there is a broader distribution for periods below \SI{10}{\day}. By constraining both of these parameters from \num{10} to \num{e5} \si{\day}, the GP parameter of the rotation period resulted in \SI{57.0\pm1.0}{\day}.

Thus far, both the spectroscopic and photometric determinations of the stellar rotation period have shown bimodal results ($\sim 28$ \si{\day} and $\sim 56$ \si{\day}), so we decided to implement the other two kernels (see \autoref{subsec:modelling}). The QPC kernel detects a rotation period of \SI{60.0\pm3.0}{\day}, while the QP kernel prefers a period of \SI{29.8\pm 1.1}{\day}. This latter result is consistent as this kernel is missing the second component at half of the rotation period (${P_\text{rot}/2}$), which is a mode of the periodic variability that often appears for activity due to the fact that any spot distribution can be to first order approximated by two spots on the star \citep{Jeffers2009}. 

In  \autoref{fig:dQ-Prot_GJ724}, we show the GP  hyper-parameters ($dQ$ and $Q$) of the dSHO kernel against the periodic component, indicating that the likelihood and number of posterior samples favour a rotation period of \SI{\sim57}{\day} instead of its second harmonic.  It is therefore reasonable to adopt the value of the stellar rotation period from the GP model with the (constrained) dSHO kernel: \SI{57.0\pm1.0}{\day}, but we cannot confidently rule out the $\sim 28$ \si{\day} as the actual rotation period. We note that the uncertainties we obtain are small, for which we interpret our result as the mean rotation period over all latitudes. Therefore, the calculated errors correspond to the errors in the mean and not to the uncertainties in the actual stellar rotation period. 

\subsection{GJ~3988}\label{sec:rotGJ3988}

\subsubsection{Spectroscopic rotation period}

\begin{figure}[!ht]
    \centering
    \includegraphics[width=\columnwidth]{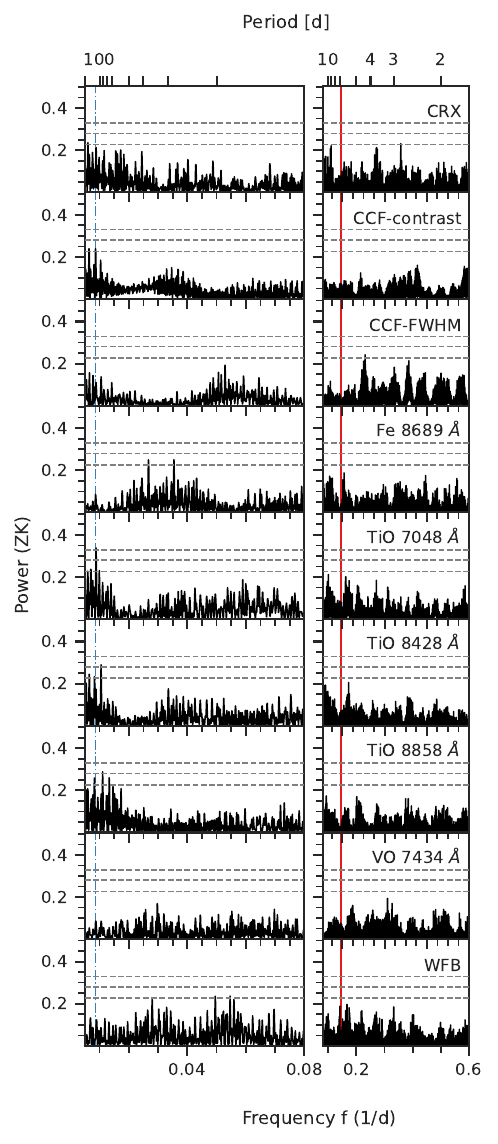} 
    \caption{Same as Fig.~\ref{fig:activity_GLS_GJ724} but for GJ~3988. The mean rotation period of \SI{116}{\day} determined from the photometry is marked by the blue dashed line. The period of the 6.4-day planet is highlighted by the red solid line.}
    \label{fig:activity_GLS_GJ3988}
\end{figure}

There are only a few significant signals in the GLS periodograms of the activity indicators obtained for GJ~3988 (see \autoref{fig:activity_GLS_GJ3988}). The strongest signal is apparent in the TiO\,7048\AA\, band index with a period of \SI{\sim114}{\day} and a $\text{FAP} < \SI{0.1}{\percent}$. Equivalently, the TiO\,9960\,{\AA} and TiO\,8428\,{\AA} indices show strong signals at \SI{\sim95}{\day} and \SI{\sim90}{\day}, which can be related to the \SI{\sim114}{\day} signal through aliasing caused by a roughly yearly sampling frequency apparent in the window function of the CARMENES observations. However, both are also close to the fourth harmonic of one year and thus might be related to the contamination of our spectra. The periodicity of \SI{115}{\day} further occurs in the periodogram of the CCF contrast, although it is only the second-strongest peak after \SI{156}{\day}, which might also be related due to aliasing.

The Fe\,8689\,\AA\, and WFB indices show less significant signals between \SI{20}{\day} and \SI{35}{\day} (FAP levels of 10\%), probably related to the moon cycle. Signals of about \SI{166}{\day} and \SI{10}{\day}, just below the \SI{10}{\percent} FAP level, are apparent in the CRX of the CARMENES VIS channel and the VO\,7434\,\AA\, band index. Considering a yearly sampling frequency, the \SI{166}{\day} period is an exact alias of the 114-day period present in the other indicators.

Since the signals of the TiO indices are the strongest in the GLS periodograms, we used a GP fit in the next step to determine a more precise estimate for the stellar rotation period from them. All three indices originate from the same molecule, which is why we performed a joint fit in which we shared the GP's rotation period between the indices. In doing so, we set the prior for the period uniform between \SI{50}{\day} and \SI{150}{\day} to cover the range of the signals visible in the GLS. The result is a spectroscopic stellar rotation period of \SI{120\pm10}{\day}.

\subsubsection{Photometric rotation period}

We used the available MEarth, SuperWasp and TJO photometry to determine the photometric rotation period of GJ~3988 as described in Section~\autoref{subsubsec:Prot-GJ724-phot} for GJ~724. A first fit was implemented using the dSHO kernel with a shared rotation period varying freely between \SI{10}{\day} and \SI{200}{\day}. The results yielded to a split posterior distribution for the rotation period, indicating an unstable signal with a low-quality factor and long-term periodicity with a period larger than \SI{\sim165}{\day} (peaked at \SI{\sim175}{\day}), along with a more stable signal at \SI{\sim115}{\day}.  Since the 115-day signal matches the period determined from the TiO band indices, we performed a second fit with constrained priors on the  difference in quality factor (\numrange[range-phrase=\,--\,]{10}{1e10}), from which we determined a photometric stellar rotation period of \SI{116\pm3}{\day}. As this determination is robust and straightforward, we did not incorporate other kernels like in the case of GJ~724. Moreover, we also interpret the small errors as the uncertainties over the mean value of the rotation period of the star. 


\section{RV data analysis}
\label{sec:analysis}

\subsection{GJ~724}
\label{subsec:GJ724}
\subsubsection{RV signal search}

\begin{figure*}
    \centering
    \includegraphics{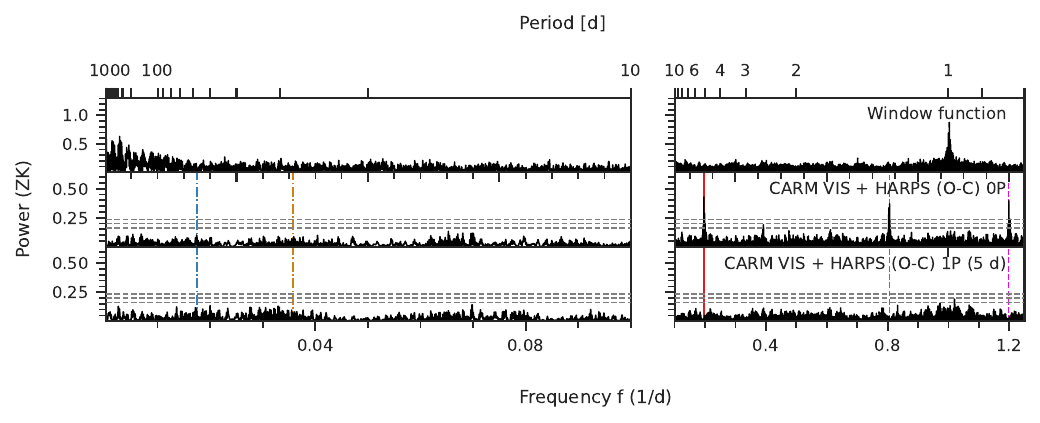} 
    \caption{GLS periodograms of the window function (top panel), RVs (middle), and residual for the one planet model (bottom) for GJ~724 using CARMENES and HARPS data sets. The period of the planet, GJ~724\,b, is highlighted by the solid red line, whereas its aliases at \SI{1.24}{\day} and \SI{0.83}{\day} are marked with dashed green and magenta lines, respectively. Additionally, we depict the mean rotation period of \SI{57}{\day} determined from the photometry and its harmonic $P_\text{rot/2}$ by the dot-dashed blue and orange lines, respectively.}
    \label{fig:gls_rv_GJ724}
\end{figure*}

\begin{figure*}
    \centering
    \includegraphics[width=0.33\textwidth]{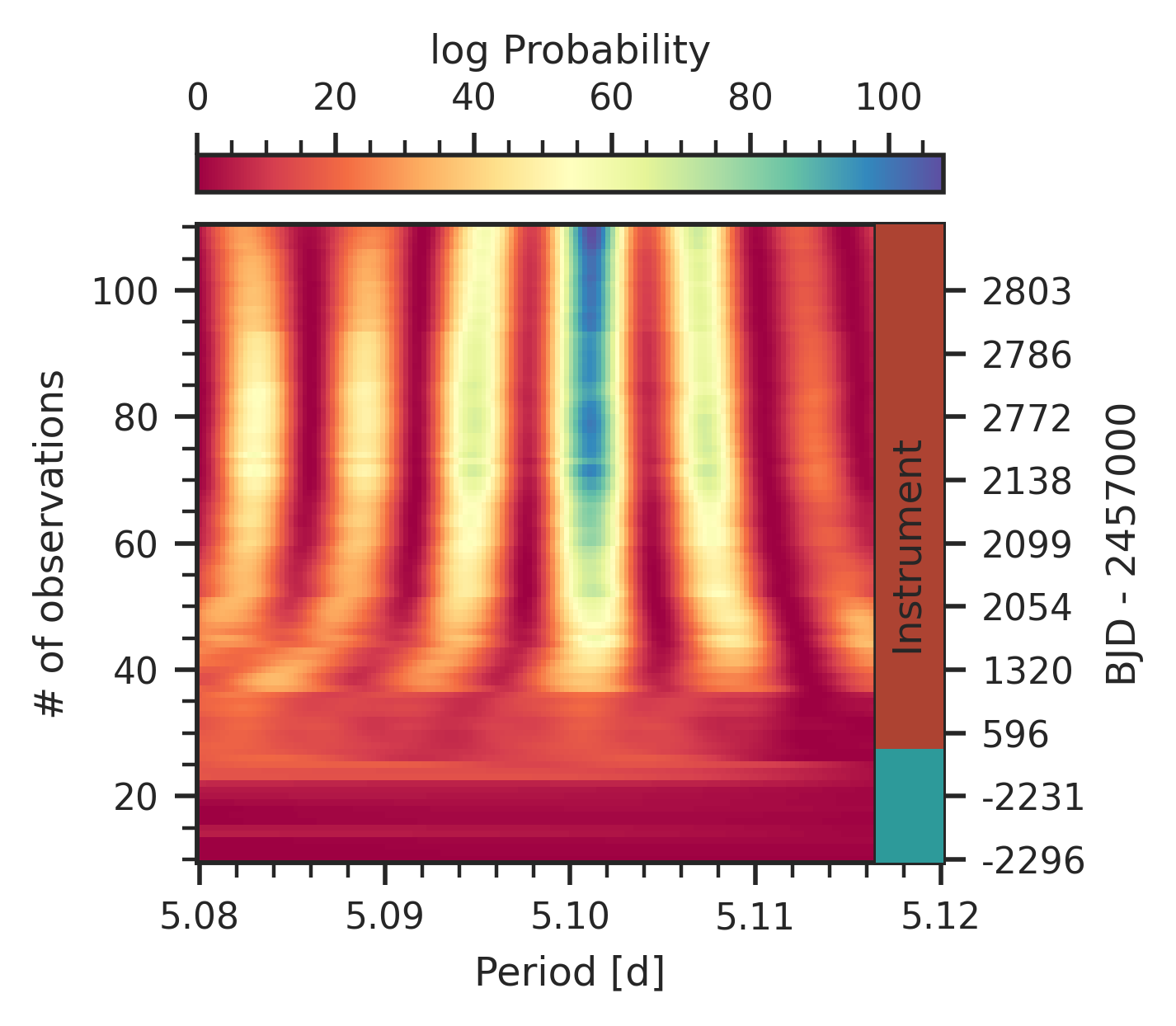}
        \includegraphics[width=0.33\textwidth]{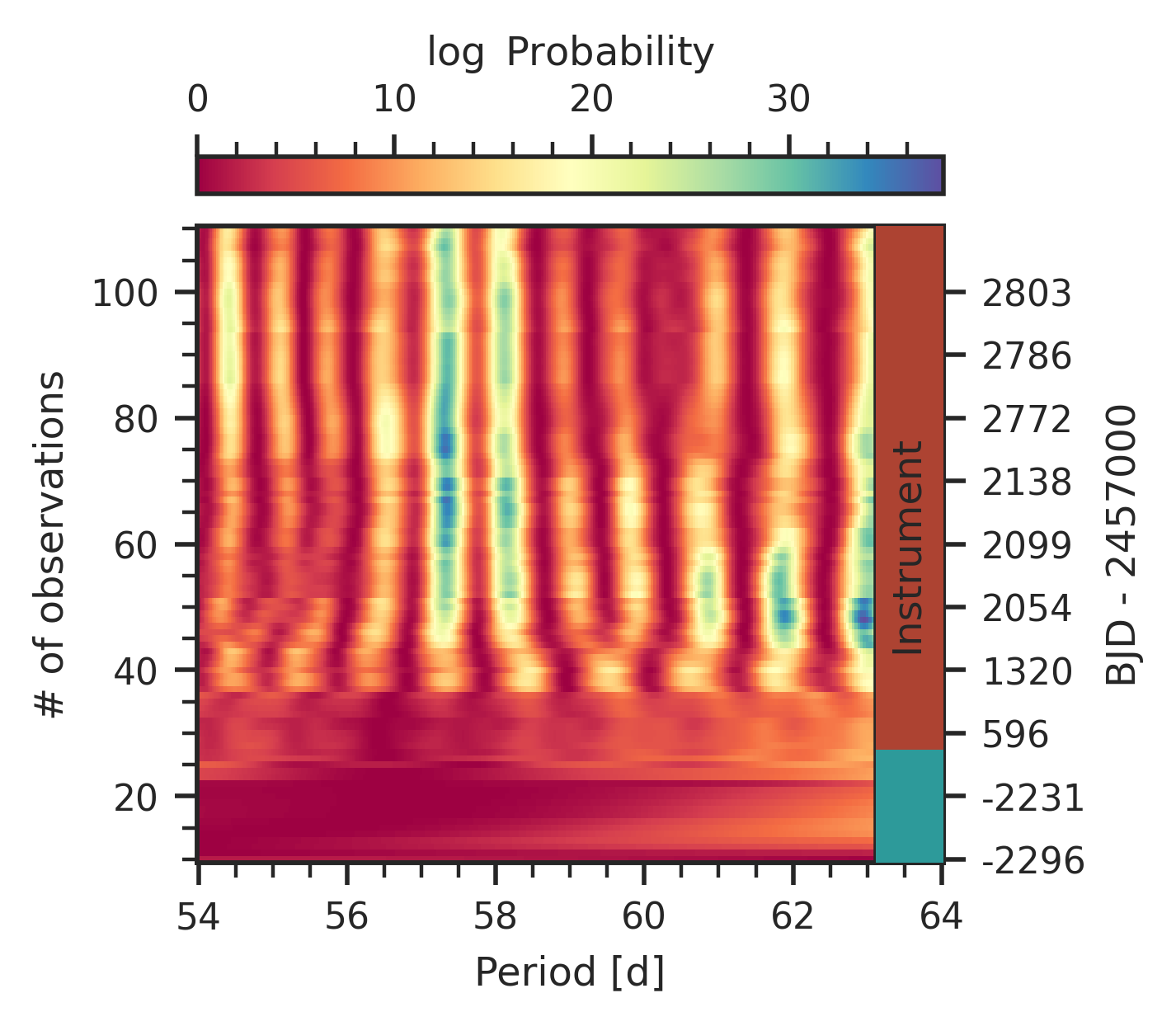}
        \includegraphics[width=0.33\textwidth]{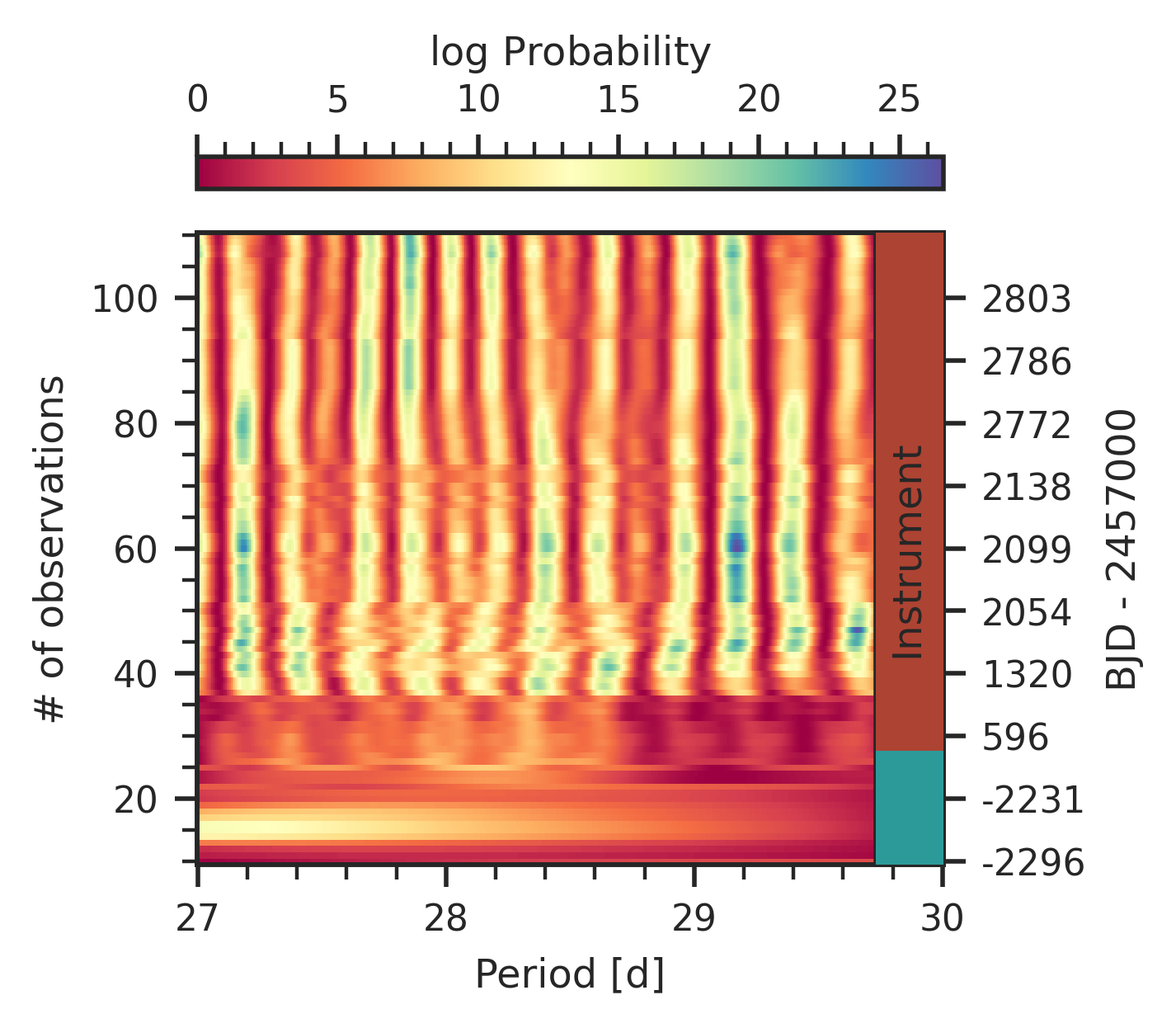}
    \caption{{\tt s-BGLS} periodogram of the RV timeseries of GJ~724. The number of data points is plotted against the period. The top colour bar indicates the logarithm of the probability, whereas the side instrument bar shows from which instrument the data originate. CARMENES is shown in orange-red and HARPS in bluish-green. The left panel is centred on the \SI{5.1}{\day} signal, middle panel around \SI{56}{\day}, and right panel around \SI{28}{\day}.}
    \label{fig:sBGLS_GJ724b}
\end{figure*}

The most significant signal present in the GLS periodogram of the RVs is at \SI{5.1}{\day} ($\text{FAP} = \SI{6.0e-7}{\percent}$), as seen in  \autoref{fig:gls_rv_GJ724}. There are also two other significant signals ($\text{FAP}$ of the order of \SIrange[range-phrase=\,--\,]{e-5}{e-4}{\percent}) with short periods ($P=\SI{0.83}{\day}$ and $P=\SI{1.24}{\day}$), due to the \SI{1}{\day} alias. Using the  \texttt{AliasFinder} \citep{Stock2020a,Stock2020b} we found \SI{5.1}{\day} to be most likely the underlying true period. The \SI{5.1}{\day} signal is not significant in any periodogram of the activity indicators and it is stable over time (as displayed in \autoref{fig:sBGLS_GJ724b}), unlike the mean stellar rotation period at \SI{57}{\day} or its harmonic at \SI{28}{\day}. There are no significant signal left in the periodogram after subtracting the \SI{5.1}{\day} signal with a single Keplerian model, but the eccentricity results in a high value ($e=0.577^{+0.055}_{-0.052}$).  Consequently, we ran a circular model, but the evidence clearly favours the eccentric one ($\Delta \ln \mathcal{Z} = - 21.6 $).


\subsubsection{RV modelling}
\label{subsubsec:GJ724-modelling}

As seen in the GLS periodogram (\autoref{fig:gls_rv_GJ724}), there are no other significant peaks besides the \SI{5.1}{\day} signal (and its aliases at $P<\SI{2}{\day}$).  Since the single Keplerian model results in a high eccentricity, we checked different configurations.  As two planets in 2:1 MMR can be misinterpreted as a single high-eccentricity planet \citep{Anglada-Escude2010}, we also tested a circular model with two planets in a 2:1 mean motion resonance. To prevent conflict with the priors, we use uniform priors on the period between \SI{4}{\day} and \SI{6}{\day} for the outer planet and between \SI{1.5}{\day} to \SI{3.5}{\day} for the inner planet. As stellar activity can affect RV data even though it does not appear significant in the GLS \citep[e.g.][]{Barnes2011, Boisse2011, Haywood2014}, we also included a GP component in some models in order to test this scenario. We use normal priors on the periodic component centred at the mean rotation period or at its second harmonic. The priors of the fits are displayed in \autoref{tab:priors_GJ724}.

\autoref{tab:evidence_GJ724} displays the comparison between the models based on their Bayesian evidence. For the Keplerian only fits (without a GP), the model with the best evidence corresponds to one planet in an eccentric orbit, as $\Delta \ln \mathcal{Z} > 5 $ when compared to the other models. For the models including Keplerians and a GP, the model with the best evidence is one eccentric planet with a GP fitted at half of the mean rotation period. This model is highly favoured over the same model but with the GP tuned to the mean rotation period ($\Delta \ln \mathcal{Z} > 5$). As shown in \autoref{fig:activity_GLS_GJ724}, many activity indicators displayed significant signals at \SI{28}{\day} (\autoref{subsubsec:Prot-GJ724-spec}), for which this result is reasonable. Since the best model comparison shows that the model including a GP with a period of \SI{56}{\day} is not as good as the model with the GP with a period of \SI{28}{\day}, we adhere to the latter. It is worth mentioning that the planetary parameters do not change between both models, as seen in \autoref{fig:model-comp-p1}. 

\begin{figure}
\centering
\includegraphics{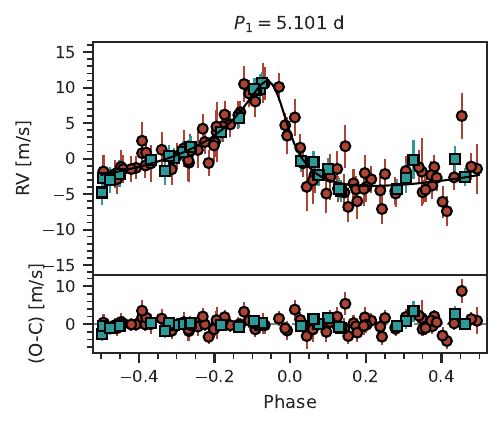}
\caption{Phased RVs for GJ~724\,b from the best fit model ($\text{1P}_\text{(5\,d-ecc)} + \text{dSHO-GP}_\text{28\,d}$). The red dots show the CARMENES data while the teal squares depicts the HARPS data. The black lines show the median of \num{10000} samples from the posterior. The residuals after subtracting the median model are shown in the lower panel.} 
\label{fig:phasefolded_GJ724}
\end{figure}

All the models favour the \SI{5.1}{\day} signal. As this is a significant signal, and there is no other counterpart in the activity indicators and photometric time series,  we conclude that it is due to a planet. We adopt the model with a GP with the highest evidence, that is the 1P$_\text{(5 d-ecc, dSHO,28d)}$ model. The RVs along this model are shown in \autoref{fig:rvs_over_time_GJ724}, while the RVs phased-folded to the planet period are shown in \autoref{fig:phasefolded_GJ724}. The posteriors and derived parameters are listed in \autoref{tab:planet_posteriors}. A corner plot of the posteriors of the planet parameters is shown in \autoref{fig:corner-planet-GJ724}.


\begin{table}
\caption{Model comparison for GJ~724.}
\label{tab:evidence_GJ724}
\begin{tabular}{l S[table-format=-3.1] S[table-format=-2.1] S[table-format=-3.1]}
\hline\hline
                         Model &  {$\ln\mathcal{Z}$} &  {$\Delta\ln\mathcal{Z}$} &  {max. $\log\mathcal{L}$} \\
\hline
\noalign{\smallskip}
    \multicolumn{4}{c}{\textit{Only Keplerian}}  \\
\noalign{\smallskip} 
               0P             &            -353.3 &                   -33.7 &        -342.9 \\
        1P$_\text{(5d-ecc)}$ &             -319.6 &                     0.0 &        -287.0 \\
       1P$_\text{(5d-circ)}$ &             -341.2 &                   -21.6 &        -315.4 \\
2P$_\text{(circ, 5d, 2.55d)}$ &            -337.3 &                   -17.7 &        -292.8 \\
\noalign{\smallskip}
    \multicolumn{4}{c}{\textit{With GP}}  \\
\noalign{\smallskip} 
           0P$_\text{(dSHO-28d)}$ &         -356.7 &  -41.4             &        -338.6 \\
 1P$_\text{(5d-ecc, dSHO, 28d)}$ &          -315.3 &   0.0              &        -269.3 \\
1P$_\text{(5d-circ, dSHO, 28d)}$ &          -342.1 &  -26.8             &        -307.9  \\
2P$_\text{(circ, 5d, 2.55d, dSHO, 28d)}$ &  -335.9 &  -20.6             &        -279.8 \\
           0P$_\text{(dSHO-56d)}$ &         -356.9 &  -41.60            &        -339.9 \\
 1P$_\text{(5d-ecc, dSHO, 56d)}$ &          -320.9 &  -5.6              &        -275.0 \\
1P$_\text{(5d-circ, dSHO, 56d)}$ &          -342.7 &  -27.4             &        -308.8 \\
2P$_\text{(circ, 5d, 2.55d, dSHO, 56d)}$ &   -339.9 &  -24.6             &        -286.0 \\
\noalign{\smallskip} 
\hline
\end{tabular}
\end{table}


\begin{table*}
    \centering
    {\setlength{\extrarowheight}{4.5pt}
        \caption{Best-fit posterior parameters and determined planet parameters for the two planetary systems.}
        \label{tab:planet_posteriors}
        \begin{tabular}{lcc}
            \hline \hline 
            Parameter\tablefootmark{(a)}                         & GJ~724 & GJ~3988      \\                               
            \hline
            \noalign{\smallskip}
            \multicolumn{3}{c}{\textit{Planetary posteriors}}                                                                        \\
            \noalign{\smallskip}
            $P_\text{b}$ (d)                                       & $\num{5.101284}^{+\num{0.000090}}_{-\num{0.000077}}$& $\num{6.9442}^{+\num{0.0010}}_{-\num{0.0010}}$     \\
            $t_{0,b}$ (BJD)                                        & $\num{2457509.766}^{+\num{0.050}}_{-\num{0.049}}$   & $\num{2457509.78}^{+\num{0.26}}_{-\num{0.25}}$        \\
            $K_\text{b}$ (\si{\meter\per\second})                  & $\num{7.48}^{+\num{0.90}}_{-\num{0.67}}$            & $\num{3.83}^{+\num{0.37}}_{-\num{0.37}}$              \\
            $\sqrt{e}_\text{b}\cos \omega_\text{b}$                & $\num{0.635}^{+\num{0.050}}_{-\num{0.055}}$         & $0$ (fixed)                                           \\
            $\sqrt{e_\text{b}}\sin \omega_\text{b}$                & $\num{0.416}^{+\num{0.060}}_{-\num{0.065}}$         & $0$ (fixed)                                           \\
            \noalign{\smallskip}
            \multicolumn{3}{c}{\textit{Derived parameters}\tablefootmark{(b)}}                                                       \\
            \noalign{\smallskip}
            $M_\text{b}\sin i$ ($M_\oplus$)                        & $\num{10.75}^{+\num{0.96}}_{-\num{0.87}}$            & $\num{3.69}^{+\num{0.42}}_{-\num{0.41}}$              \\
            $a_\text{b}$  (\si{\astronomicalunit})                 & $\num{0.04685}^{+\num{0.00077}}_{-\num{0.00079}}$    & $\num{0.0405}^{+\num{0.0011}}_{-\num{0.0012}}$        \\  
            $e_\text{b}$                                           & $\num{0.577}^{+\num{0.055}}_{-\num{0.052}}$          & $0$ (fixed)                                           \\
            $\omega_\text{b}$ (deg)                                & $\num{33.2}^{+\num{5.7}}_{-\num{5.5}}$               & $0$ (fixed)                                           \\
            $<S_\text{b}$ ($S_\oplus$)>                            &  $\num{23.32}^{+\num{0.82}}_{-\num{0.77}}$           & $\num{2.45}^{+\num{0.15}}_{-\num{0.13}}$              \\
            $<T_\text{eq, b}$\tablefootmark{({c})} (\si{\kelvin})> & $\num{611}^{+\num{20}}_{-\num{19}}$                  & $\num{348}^{+\num{17}}_{-\num{16}}$                   \\
            $T_\text{eq, b}$\tablefootmark{({d})} (\si{\kelvin})   & $\num{643}^{+\num{30}}_{-\num{26}}$                  & $\num{349}^{+\num{19}}_{-\num{17}}$                   \\

            \noalign{\smallskip}
            \multicolumn{3}{c}{\textit{GP posteriors}}                                                                               \\
            \noalign{\smallskip}
            $P_\text{GP, rv}$ (d)                                    &  $\num{29.50}^{+\num{0.88}}_{-\num{1.46}}$          & $\num{115.28}^{+\num{3.24}}_{-\num{3.06}}$           \\
            $\sigma_\text{GP, rv, CARM-VIS}$ (\si{\meter\per\second})&  $\num{3.09}^{+\num{1.14}}_{-\num{0.72}}$           & $\num{3.30}^{+\num{1.20}}_{-\num{0.85}}$             \\
            $\sigma_\text{GP, rv, HARPS}$ (\si{\meter\per\second})   &  $\num{3.00}^{+\num{1.17}}_{-\num{0.86}}$           &   \dots                                              \\
            $\sigma_\text{GP, rv, IRD}$ (\si{\meter\per\second})     &    \dots                                            & $\num{4.36}^{+\num{2.20}}_{-\num{1.50}}$             \\
            $f_\text{GP, rv}$                                        &  $\num{0.70}^{+\num{0.21}}_{-\num{0.31}}$           & $\num{0.65}^{+\num{0.24}}_{-\num{0.31}}$             \\
            $Q_{0, \text{GP, rv}}$                                   &  $\num{5.4}^{+\num{15.4}}_{-\num{4.5}}$             & $\num{0.46}^{+\num{1.10}}_{-\num{0.29}}$             \\
            $dQ_\text{GP, rv}$                                       &  $\num{23}^{+\num{3193}}_{-\num{23}}$               & $\num{561.36}^{+\num{19571.86}}_{-\num{560.55}}$        \\

            \noalign{\smallskip}
            \multicolumn{3}{c}{\textit{Instrumental posteriors}}                                                                     \\
            $\gamma_\text{CARM-VIS}$ (\si{\meter\per\second})       & $\num{0.79}^{+\num{0.40}}_{-\num{0.43}}$             & $\num{0.30}^{+\num{0.61}}_{-\num{0.59}}$             \\
            $\sigma_\text{CARM-VIS}$   (\si{\meter\per\second})     & $\num{1.18}^{+\num{0.67}}_{-\num{0.69}}$             & $\num{1.27}^{+\num{0.49}}_{-\num{0.50}}$             \\
            $\gamma_\text{HARPS}$ (\si{\meter\per\second})          & $\num{-0.80}^{+\num{0.66}}_{-\num{0.63}}$            &  \dots                                               \\
            $\sigma_\text{HARPS}$   (\si{\meter\per\second})        & $\num{0.65}^{+\num{0.70}}_{-\num{0.45}}$             &  \dots                                               \\
             $\gamma_\text{IRD}$ (\si{\meter\per\second})           & \dots                                                & $\num{-0.12}^{+\num{1.04}}_{-\num{1.50}}$         \\
            $\sigma_\text{IRD}$   (\si{\meter\per\second})          & \dots                                                & $\num{3.17}^{+\num{0.77}}_{-\num{0.72}}$             \\
            \noalign{\smallskip}
            \hline
        \end{tabular}} 
    \tablefoot{\tablefoottext{a}{Error bars denote the $68\%$ posterior credibility intervals.}
        \tablefoottext{b}{These derived parameters were computed using the stellar parameters that were sampled from a normal distribution centred around the values from \autoref{tab:stellar-properties} with the width of the uncertainty.}
        \tablefoottext{c}{Assuming a zero Bond albedo. }
        \tablefoottext{d}{Computed using Eq. (4) from \cite{Quirrenbach2022}.}
        }
\end{table*}

\subsection{GJ~3988}
\label{subsec:GJ3988}

\subsubsection{RV signal search}

\begin{figure*}
    \centering
    \includegraphics{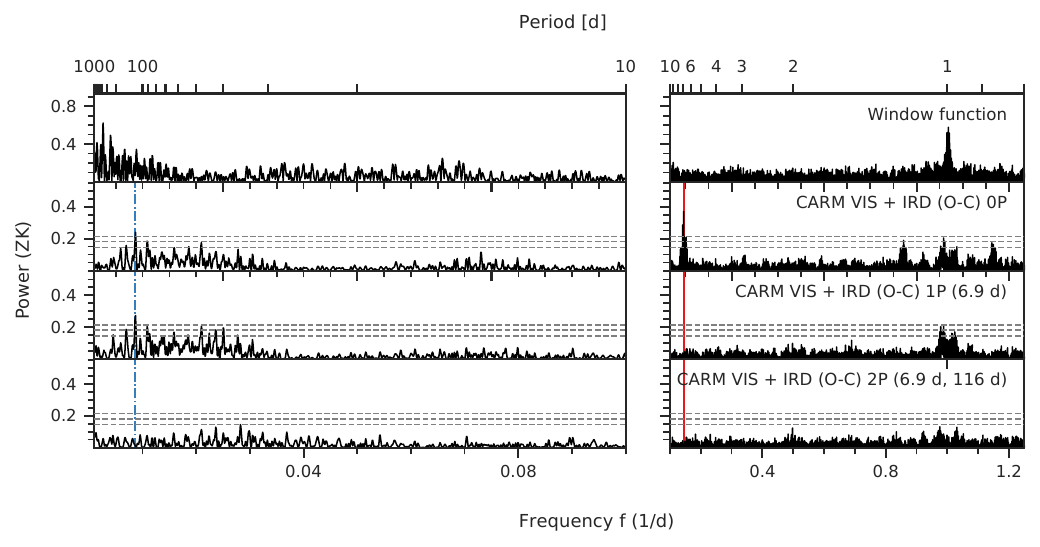} 
    \caption{GLS periodograms and window function of the CARMENES and IRD RVs for GJ~3988. The period of the planet, GJ~3988\,b, is highlighted by the solid line and the mean rotation period of \SI{116}{\day} determined from the photometry is marked by the blue dashed line.}
    \label{fig:gls_rv_GJ3988}
\end{figure*}

\begin{figure*}
    \centering
    \includegraphics[width=0.4\textwidth]{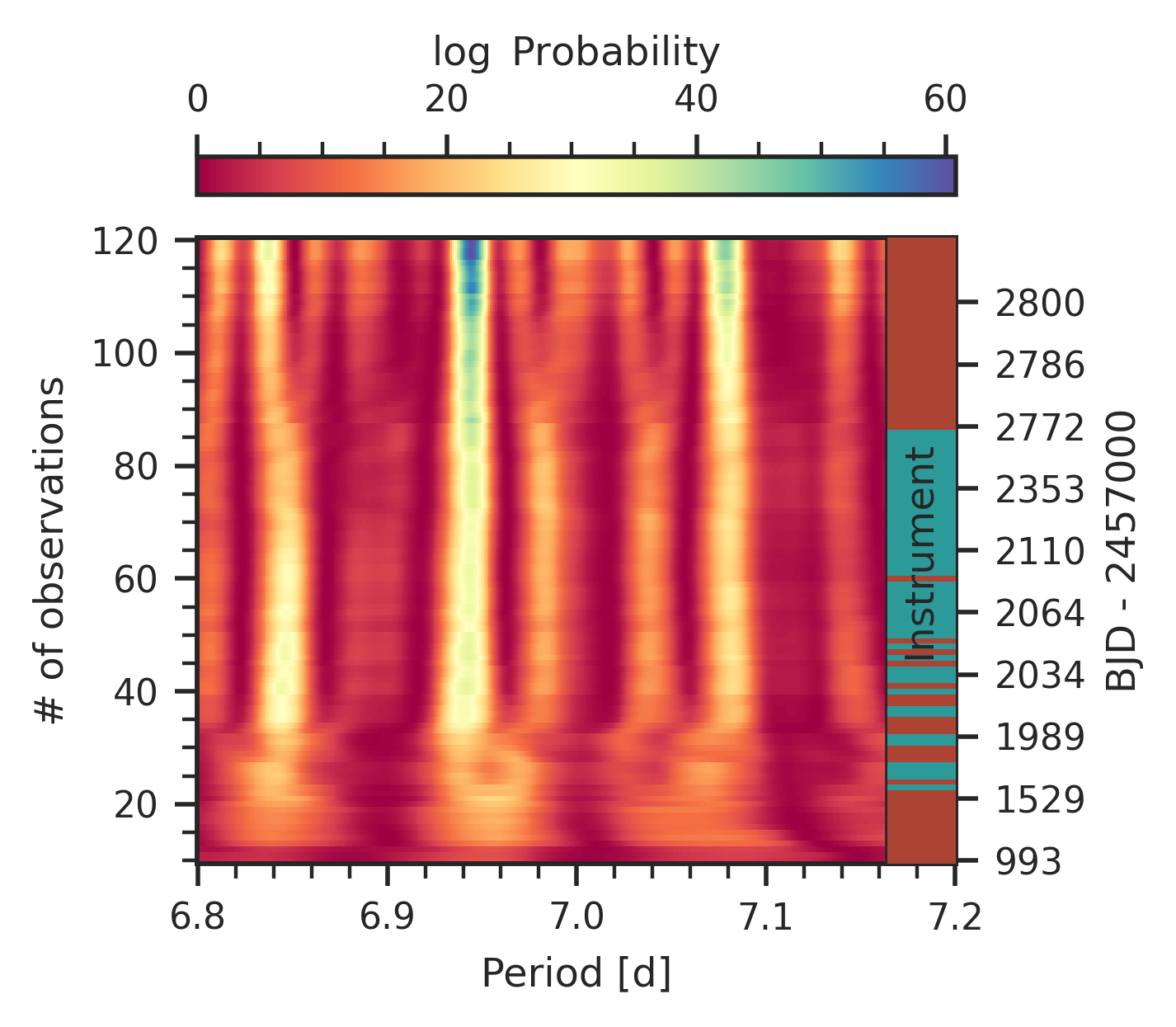}
    \includegraphics[width=0.4\textwidth]{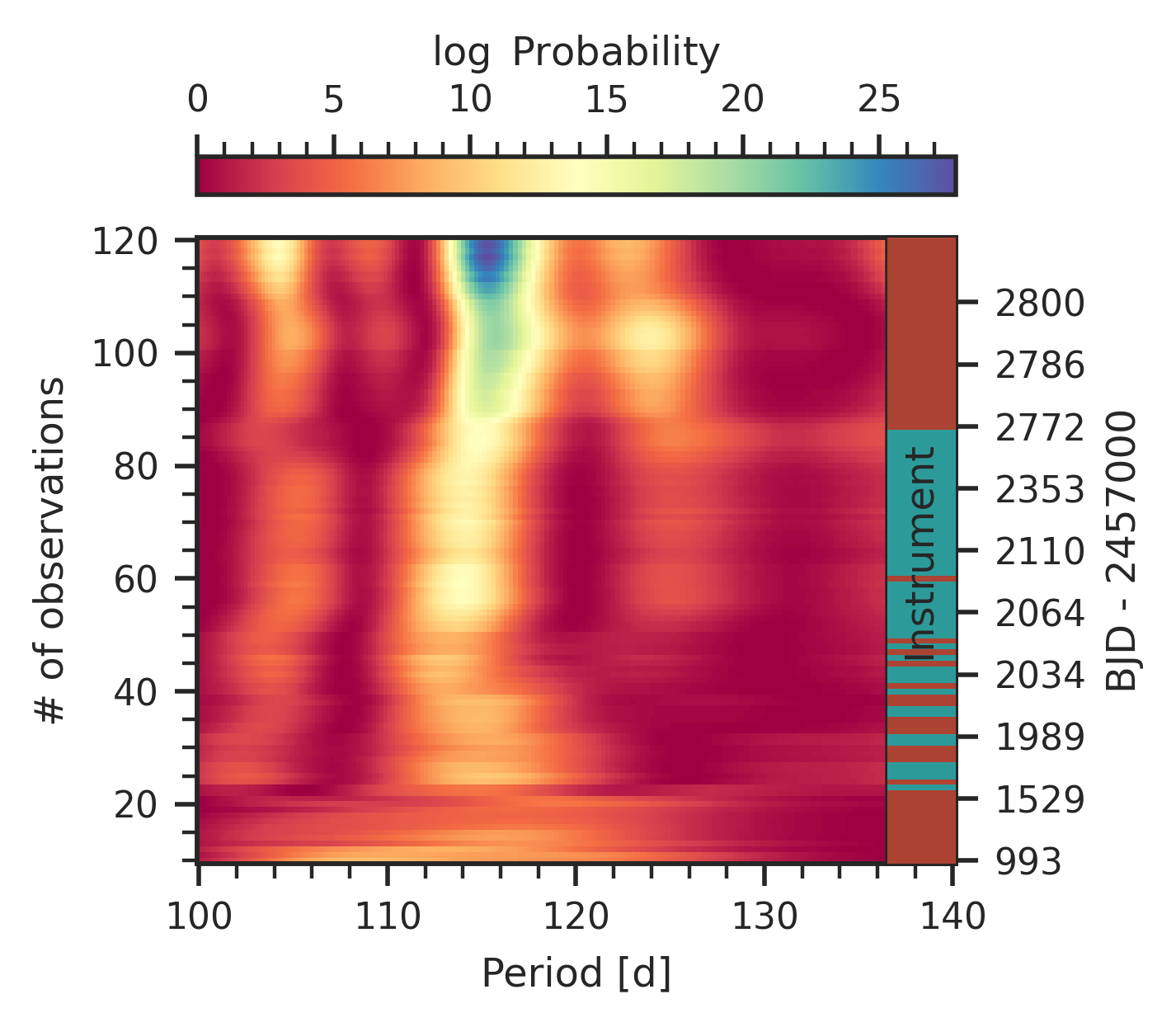}
    \caption{Same as Fig. \ref{fig:sBGLS_GJ724b} but for GJ~3988. In the instrument coloured bar, CARMENES is shown in orange-red, while IRD in bluish-green. Left panel is centred on the 6.94-day signal, after subtracting the stellar activity using a GP model. Right panel is centred on the imprint of the mean stellar rotation period after subtracting the 6.94-day signal from the data.}
    \label{fig:sBGLS_GJ3988}
\end{figure*}

The GLS periodogram of the CARMENES and IRD data for GJ~3988 (presented in \autoref{fig:gls_rv_GJ3988}) shows a highly significant signal with a period of 
\SI{6.94}{\day} ($\text{FAP} < \SI{1e-6}{\percent}$) accompanied by slightly less significant aliases, due to the dominant daily sampling frequency, at \SI{1.16}{\day} and \SI{0.87}{\day}. If the signal is subtracted from the data using a circular Keplerian model, the highest peak in the GLS periodogram is at \SI{115}{\day}, which is consistent with our measurement of GJ~3988's mean rotation period from \autoref{sec:rotGJ3988}. The large number of other GLS peaks surrounding the 115-day signal suggests that this is indeed an imprint of the stellar activity. The residuals of a simultaneous fit to the 6.94-day and 115-day signals do not show any further significant signals. The highest remaining peak at \SI{35.35}{\day} has a FAP of about \SI{70}{\percent}.

The {\tt s-BGLS} of the 6.94-day signal, as a measure of its stability over time, is shown in the left panel of \autoref{fig:sBGLS_GJ3988}. We do not detect any variability in the signal but only a steady increase in probability, which suggests a static nature for it. On the other hand, the {\tt s-BGLS} focused on \SI{115}{\day} (right panel) shows slight variations in the probability over time in combination with shifts of the period, which indicates a signal of only quasi-periodic nature, as we would expect for signals related to stellar activity.

\subsubsection{RV modelling}

In the previous section, we detected two significant signals in the RVs of GJ~3988: a planet candidate signal with a period of \SI{6.94}{\day} and an imprint of the mean stellar rotation period at \SI{\sim116}{\day}. We consequently compared a model considering pure stellar activity against models that additionally included the planet candidate. In doing so, we considered both, circular and eccentric orbits. The priors of the fits are listed in \autoref{tab:priors_GJ3988}.

\begin{table}
\caption{Model comparison for GJ~3988.}
\label{tab:evidence_GJ3988}
\begin{tabular}{l S[table-format=-3.1] S[table-format=-2.1] S[table-format=-3.1]}
\hline\hline
                         Model &  {$\ln\mathcal{Z}$} &  {$\Delta\ln\mathcal{Z}$} &  {max. $\log\mathcal{L}$} \\
\hline
\noalign{\smallskip}
    \multicolumn{4}{c}{\textit{Only Kepler}}  \\
\noalign{\smallskip} 
                    0P             & -379.3            &    -15.9                &        -368.2 \\
             1P$_\text{(7 d-ecc)}$ & -373.9            &    -10.5                &        -342.2 \\
            1P$_\text{(7 d-circ)}$ & -374.8            &    -11.4                &        -342.5 \\
 2P$_\text{(7 d-ecc, 116 d-circ)}$ & -368.4            &    -5.0                 &        -323.1  \\
2P$_\text{(7 d-circ, 116 d-circ)}$ & -363.4            &    -0.0                 &        -324.2  \\
\noalign{\smallskip}
    \multicolumn{4}{c}{\textit{With GP}}  \\
\noalign{\smallskip} 
        0P + dSHO-GP$_\text{(116 d)}$             & -378.5            & -17.4     &       -358.8 \\
1P$_\text{(7 d-circ)}$ + dSHO-GP$_\text{(116 d)}$ & -361.1            & -0.0      &       -324.0 \\
 1P$_\text{(7 d-ecc)}$ + dSHO-GP$_\text{(116 d)}$ & -372.2            & -11.1     &       -323.1 \\
\hline
\end{tabular}
\end{table}

The results of the model comparison using the Bayesian evidence are presented in \autoref{tab:evidence_GJ3988}. We found that the models including the planet candidate are highly significant compared to the activity-only model. Based on this, in combination with the results from the analysis of the activity indicators and the {\tt s-BGLS}, we qualified the signal as a genuine planetary signal. For the 'only Keplerian' models, the best one  is the two-Keplerian in circular orbit, which corresponds to the planet plus a static sinusoidal component used to model the stellar activity ($P_\text{rot}$=\SI{116}{\day}). Moreover, for the models with Keplerian plus GP, the model with one planet in a circular orbit plus a GP is strongly preferred ($\Delta \mathcal{\ln{Z}} > 5.0$). Assuming that the circular model for the planet in combination with the quasi-periodic GP for the stellar activity are better physically motivated compared to two sinusoidal components, we chose the circular $\text{1P}_\text{(7\,d-circ)} + \text{dSHO-GP}_\text{(116\,d)}$ as our best fit model.



\begin{figure}
\centering
\includegraphics{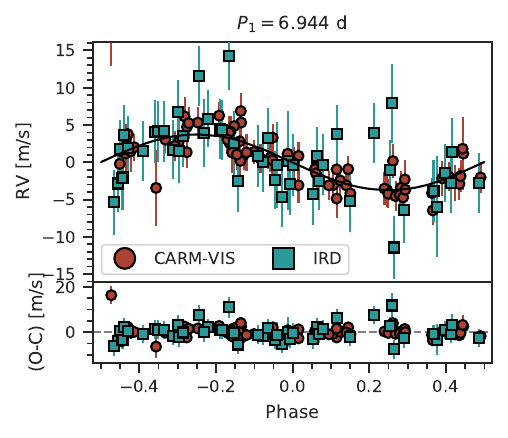}
\caption{Same as \autoref{fig:phasefolded_GJ724} but for the best-model for GJ~3988\,b ($\text{1P}_\text{(7\,d-circ)} + \text{dSHO-GP}_\text{116\,d}$).  The red dots show the CARMENES data while the teal squares depicts the IRD data.} 
\label{fig:phasefolded_GJ3988}
\end{figure}

The posteriors from this fit are listed in \autoref{tab:planet_posteriors} together with the derived parameters for the planet. In \autoref{fig:rvs_over_time_GJ3899} and \autoref{fig:phasefolded_GJ3988}, we show the RVs together with the best-fit model and the phase-folded plot of the planet, respectively.


\subsection{Transit search and detection limits}
In this section, we aim to confirm or refute the transiting nature of the planets GJ~724\,b and GJ~3988\,b, by exploring the public data provided by the TESS mission \citep{ricker2015}. On the one hand, GJ~724\,b has not been observed yet, but will be in sector 80 during the second TESS extended mission. Moreover, the current ephemeris for this planet provides observational windows with an uncertainty of about 6~h, which makes any time-critical observation using ground-based facilities very challenging. Then, unfortunately, we can not reach any conclusion concerning its transiting nature. On the other hand, GJ~3988\,b has been observed in multiple sectors and cadences: during the primary mission in sectors 24, 25, and 26 using the 2~min and 30~min cadences, and during the extended mission  in sectors 50 and 51 using the 20~sec, 2~min and 10~min cadences, and in sectors 52 and 53 using the 20~sec and the 2~min. For our study, we used all the data corresponding to the 2~min cadence, which cover seven sectors in total.

Neither Science Processing Operations Center \citep[SPOC;][]{Jenkins2016} nor the Quick-Look Pipeline \citep[QLP; ][]{Huang2020} have issued an alert for GJ~3988, which is a hint of the non-transiting nature of this system. However, these pipelines may not detect some periodic transits if they are shallow and their S/N is below their detection thresholds. Then, the community is often conducting complementary planetary searches either using ground-based facilities \citep[see, e.g.,][]{Delrez2022} 
or using alternative custom detection pipelines. In this context, we explored the TESS data using the \sherlock pipeline\footnote{\sherlock ({S}earching for {H}ints of {E}xoplanets f{R}om {L}ightcurves {O}f spa{C}e-based see{K}ers) code is fully available on GitHub: \url{https://github.com/franpoz/SHERLOCK}}, presented initially by \cite{Pozuelos2020} and \cite{Demory2020}, and used in several studies \citep[see, e.g.,][]{Wells2021,VanGrootel2021,Schanche2022}. This pipeline allows the user to explore TESS data to recover known planets and alerts and to search for new periodic signals, which may hint at the existence of extra-transiting planets. In short, the pipeline combines six modules to (1) download and prepare the light curves from the MAST, (2) search for planetary candidates, (3) perform a semi-automatic vetting of the interesting signals, (4) compute a statistical validation, (5) model the signals to refine their ephemerides, and (6) compute observational windows from ground-based observatories to trigger a follow-up campaign. We refer the reader to \cite{Delrez2022} and \cite{Pozuelos2023} for recent \sherlock applications and further details. 

During our search, we explored an orbital range from 0.5 to 40~d, employing a density grid of 247260 periods distributed in a logarithmic distribution, with shorter time steps for shorter orbital periods and larger time steps for longer orbital periods. The optimum period grid to explore was computed based on the stellar mass and radius and the time span covered by the data set \citep{Hippke2019}. In our case, the minimum and maximum time steps were $4.65\times10^{-6}$ and \SI{0.0016}{\day}, respectively. A minimum of two transits was required to claim a detection.

We did not find any hint of a periodic signal which might correspond to GJ~3988\,b or any other transiting planet in the system. All the signals detected were attributable to systematics or noise. Then, in this scenario, we wondered if the photometric precision provided by TESS is enough to detect the presence of GJ~3988\,b if we assume that it is a transiting planet. To answer this question, we conducted an injection-and-recovery experiment using the \matrixtk\footnote{{The \matrixtk ({M}ulti-ph{A}se {T}ransits {R}ecovery from {I}njected e{X}oplanets) code is open access on GitHub: \url{https://github.com/PlanetHunters/tkmatrix}}} code \citep{Devora-Pajares2022}. \matrixtk injects synthetic planets over the 2~min cadence light curves corresponding to the seven sectors used in this study.  

If we assume that GJ~3988\,b is a transiting planet, employing the mass-radius relationship given by \cite{Chen2017}, its radius would be 1.72$^{+0.5}_{-0.7}$~R$_{\oplus}$. Then, we explored the $R_{\mathrm{planet}}$--$P_{\mathrm{planet}}$ parameter space in the ranges of 0.5--3.0\,R$_{\oplus}$ with steps of 0.13\,R$_{\oplus}$, and 0.5-8.0 days with steps of 0.20 days. Moreover, for each combination of $R_{\mathrm{planet}}$--$P_{\mathrm{planet}}$ \matrixtk explores three different phases, that is, different values of $T_{0}$. In total, we explored 1350 scenarios. 
Once the synthetic planets are injected, \matrixtk detrends the light curves using a bi-weight filter with a window size of 0.5~d, which was the optimal value during the \sherlock search. \matrixtk considers a synthetic planet as recovered when its epoch matches the injected epoch with 1~hour accuracy, and its period is within 5\,\% of the injected period. 

It is worth noting that for simplicity, the injected planets have impact parameters and eccentricities equal to zero and that since we injected the synthetic signals in the PDCSAP light curve, these signals were not affected by the PDCSAP systematic corrections; hence, the detection limits that we find can be considered as the most optimistic scenario \citep[see e.g.][]{Pozuelos2020,Eisner2020}.

\begin{figure}
\includegraphics[width=\columnwidth]{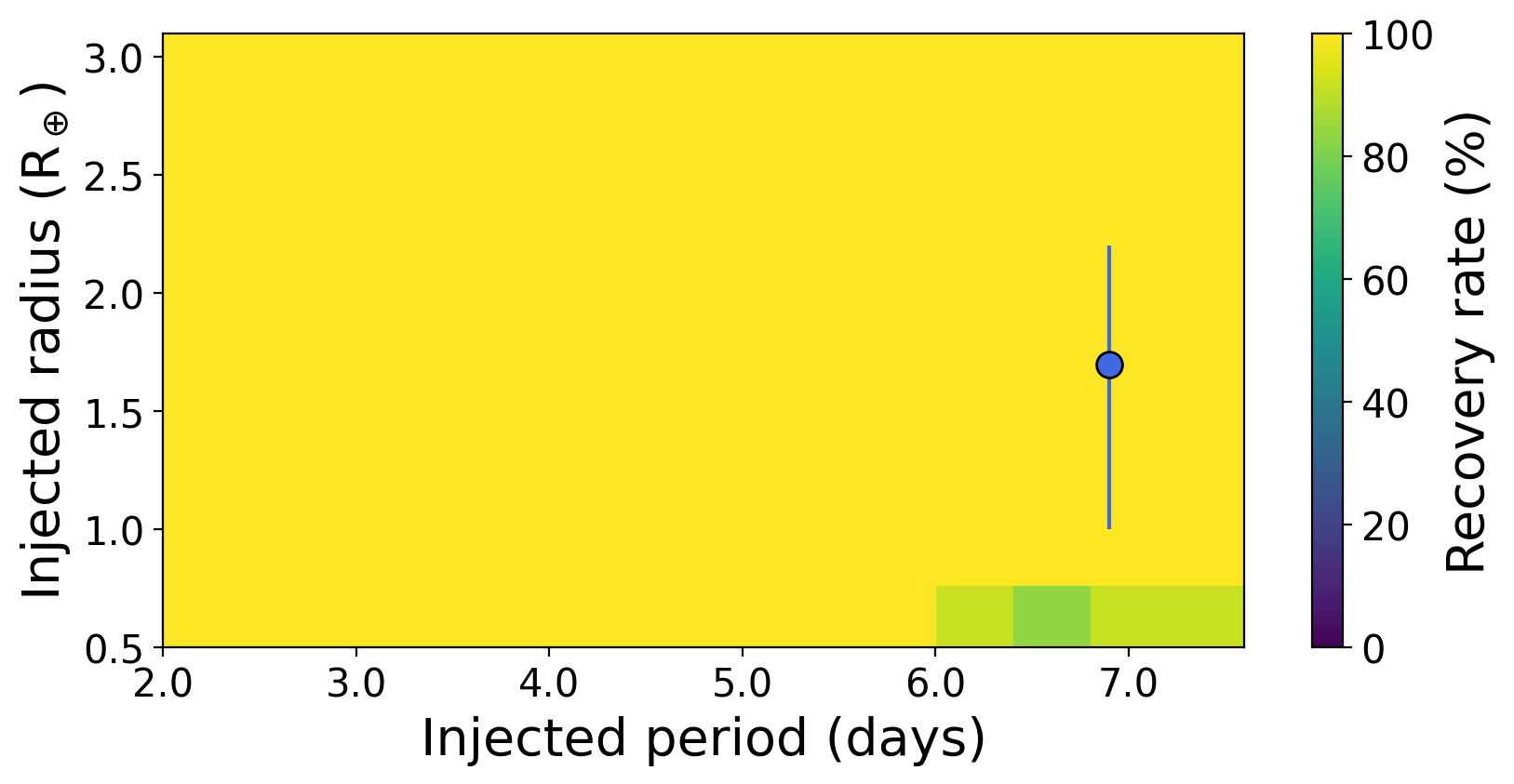}
\caption{Detectability map of GJ 3988\,b using the TESS 2~min data corresponding to sectors 24, 25, 26, 50, 51, 52, and 53. We explored a total of 1350 different scenarios. Larger recovery rates are presented in yellow and green colours, while lower recovery rates are in blue and darker hues. In the parameters space explored, each combination of $R_{\mathrm{planet}}$--$P_{\mathrm{planet}}$ yielded a recovery rate of nearly 100\%, which allows us to confirm the non-transiting nature of this system robustly. The blue dot refers to the planet GJ 3988\,b assuming its transiting nature.} \label{fig:recovery}
\end{figure}

The detectability map resulting from this injection-and-recovery experiment is shown in Fig.~\ref{fig:recovery}. We found that any planet within the range of the orbital period and radius explored in this experiment would be easily detected, yielding a recovery rate of nearly 100\%. Hence, we can robustly confirm the non-transiting nature of GJ~3988\,b, and the non-existence of any other transiting planet with an orbital period shorter than 8~d and radius larger than 0.5~R$_{\oplus}$.  

\section{Discussion}
\label{sec:discussion}

\subsection{GJ~724}

\begin{figure}
    \centering
    \includegraphics[width=\hsize]{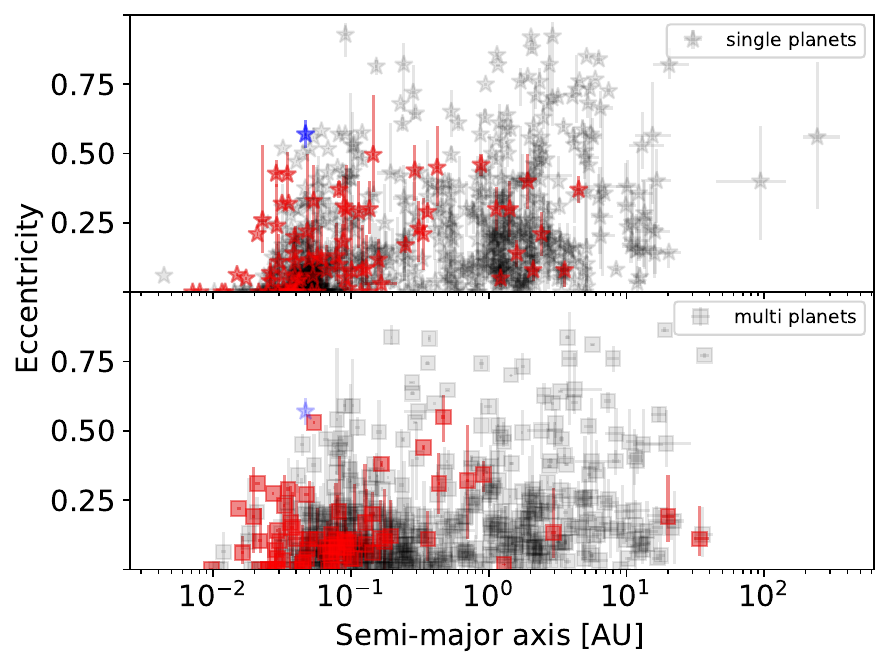}
    \caption{Eccentricity plotted against semi-major axis of the confirmed exoplanets around single stars from the
    \href{https://exoplanetarchive.ipac.caltech.edu/}{NASA Exoplanet Archive}. The red colour represent planets around M-dwarfs, whereas the gray colour illustrates planets around any other stellar mass. Single planets (top panel) are represented by a star symbol, while planets in multi-planetary systems are represented by a square symbol (bottom). The blue star shows the position of GJ~724 b from the solution of the single planet in an eccentric orbit, which is also plotted faintly in the bottom panel (multi-planetary systems) just for visualisation purposes.}
    \label{fig:GJ724_ecc_a_diagram}
\end{figure}

GJ~724\,b has a very high eccentricity ($e \sim 0.6$). We  tested the orbital configuration of two planets in a circular 2:1 MMR, with the eccentric model being the preferred one ($\Delta\ln\mathcal{Z} > 5$). In \autoref{fig:GJ724_ecc_a_diagram} we plot GJ~724\,b alongside all confirmed exoplanets around single stars, observing that this planet is the most eccentric planet amongst those with proximity to the star ($a \le 0.05$ \si{\astronomicalunit}) and is the most eccentric of all planets around M dwarfs.
There are other systems with single, high-eccentricity ($e > 0.3$) planets around M dwarfs, such as GJ~4276\,b ($M_p\sin{i} = 16.57^{+0.94}_{-0.95}$\,$M_\oplus$, $a$ $= 0.082 \pm 0.002$\,au, $P = 13.352 \pm 0.003$\,d, $e=0.37 \pm 0.03$; \citealt{Nagel2019}), GJ 514\,b ($M_p\sin{i}= 5.2 \pm 0.9$\,$M_\oplus$, $a$ $= 0.422^{+0.014}_{-0.015}$\,au, $P =140.43 \pm 0.41$\,d, $e= 0.45^{+0.15}_{-0.14} $; \citealt{Damasso2022}), and GJ~96\,b ($M_p\sin{i}$ $= 19.66^{+2.42}_{-2.30}$\,$M_\oplus$, $a$ $= 0.291\pm 0.005$\,au, $P = 73.94^{+0.33}_{-0.38}\,d$, and $e= 0.44^{+0.09}_{-0.11} $; \citealt{Hobson2018}).

\begin{figure}
    \centering
    \includegraphics[width=\hsize]{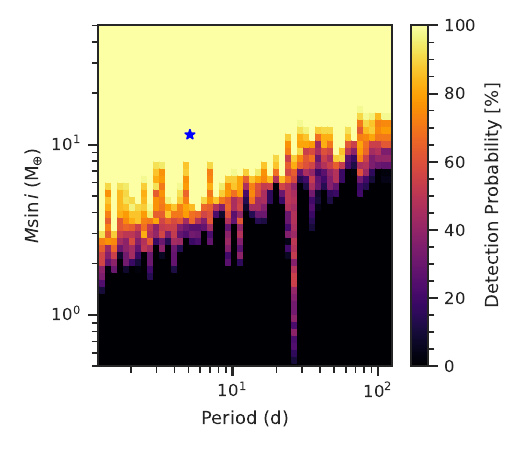}
    \caption{Detection limit map of the RVs of GJ~724. The colour map represents the detection probability of the period-mass combination. The Grid is $50 \times 50$ in mass and period, with 50 phase samples for each combination. The blue star indicates the planet GJ~724\,b. }
    \label{fig:GJ724_detection_map}
\end{figure}

One explanation for the existence of close planets in eccentric orbits is the presence of  another companion in the system. By analysing the periodiogram of the RVs residuals of GJ~724, we do not find any signal that could be attributed to a second planet. If there is another companion, it is more likely that it has a longer orbital period, perturbing the orbit of the innermost planet by gravitational interaction and therefore explaining the high eccentricity of GJ~724\,b.  With the aim to check if there is a possible planet that we are not able to detect, we computed a detection limit map using the same method as in \citet{Sabotta2021} (displayed in \autoref{fig:GJ724_detection_map}). GJ~724\,b lies in the region with 100$\%$ detection probability. A possible companion with a larger minimum mass could have been detected up to an orbital period of \SI{\sim100}{\day}. Any potential disturber must have a lower minimum planetary mass than GJ~724\,b in order to escape detection with our dataset. 

\begin{figure}
        \includegraphics[width=\hsize]{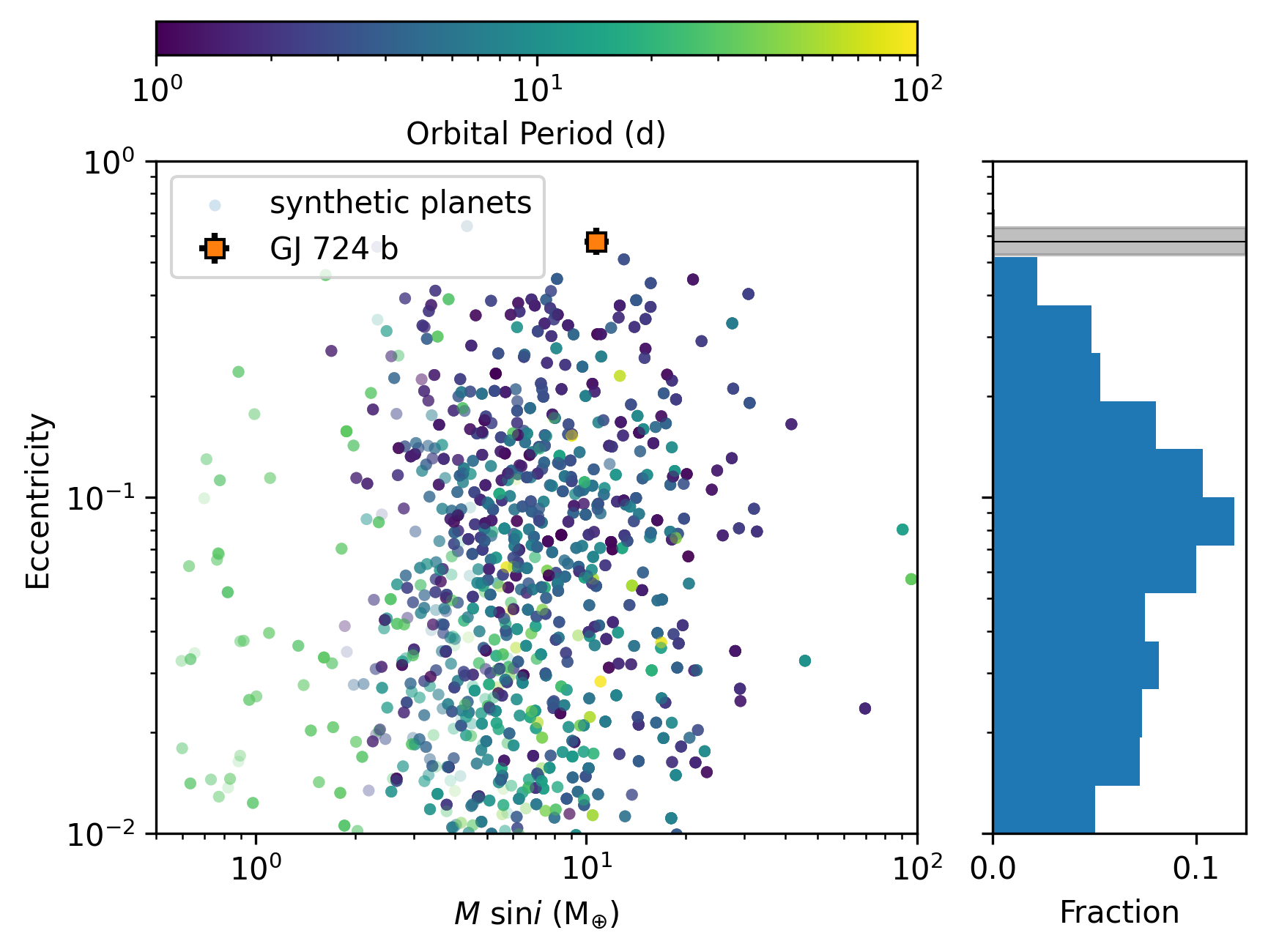}
        \caption{Synthetic planetary population as a function of $M \sin i$ and eccentricity. The colour shows the orbital period of the planets and the histogram depicts the fraction of theoretically expected observed planets, which is normalised by the number of planets in the same sample.The grey region corresponds to the eccentricity of GJ~724\,b within $1 \sigma$ range. }
        \label{fig:synthetic_PMsiniEcc}
\end{figure}

The population of synthetic planets orbiting a 0.5\,$M_{\odot}$ star, as studied by \citet{Burn2021}, can be used to investigate the prevalence of such systems as predicted by conventional planet formation models \citep{Emsenhuber2021}. We proceed by randomly drawing multiple inclinations for each system and applying the detection sensitivity for GJ 724 shown in \autoref{fig:GJ724_detection_map}. In this manner, synthetic planet detections can be generated with repeated inclusion of individual synthetic planets for improved statistical significance. This approach is described in detail in \citet{Schlecker2022}. In \autoref{fig:synthetic_PMsiniEcc} we show the resulting synthetic population of planets and their eccentricities resulting from theoretical N-body integration over \SI{20}{\mega \year} and individual planet evolution including mass-loss and tidal in-spiral of \SI{5}{\giga \year}. It was applied to the synthetic planets with assigned inclinations drawn from an isotropically uniform distribution. We can see that the model generally fails to reproduce GJ~724\,b-like planets. However, those eccentricity values do not include long-term tidal damping, which is discussed later in this paper.

\begin{figure}
        \includegraphics[width=\hsize]{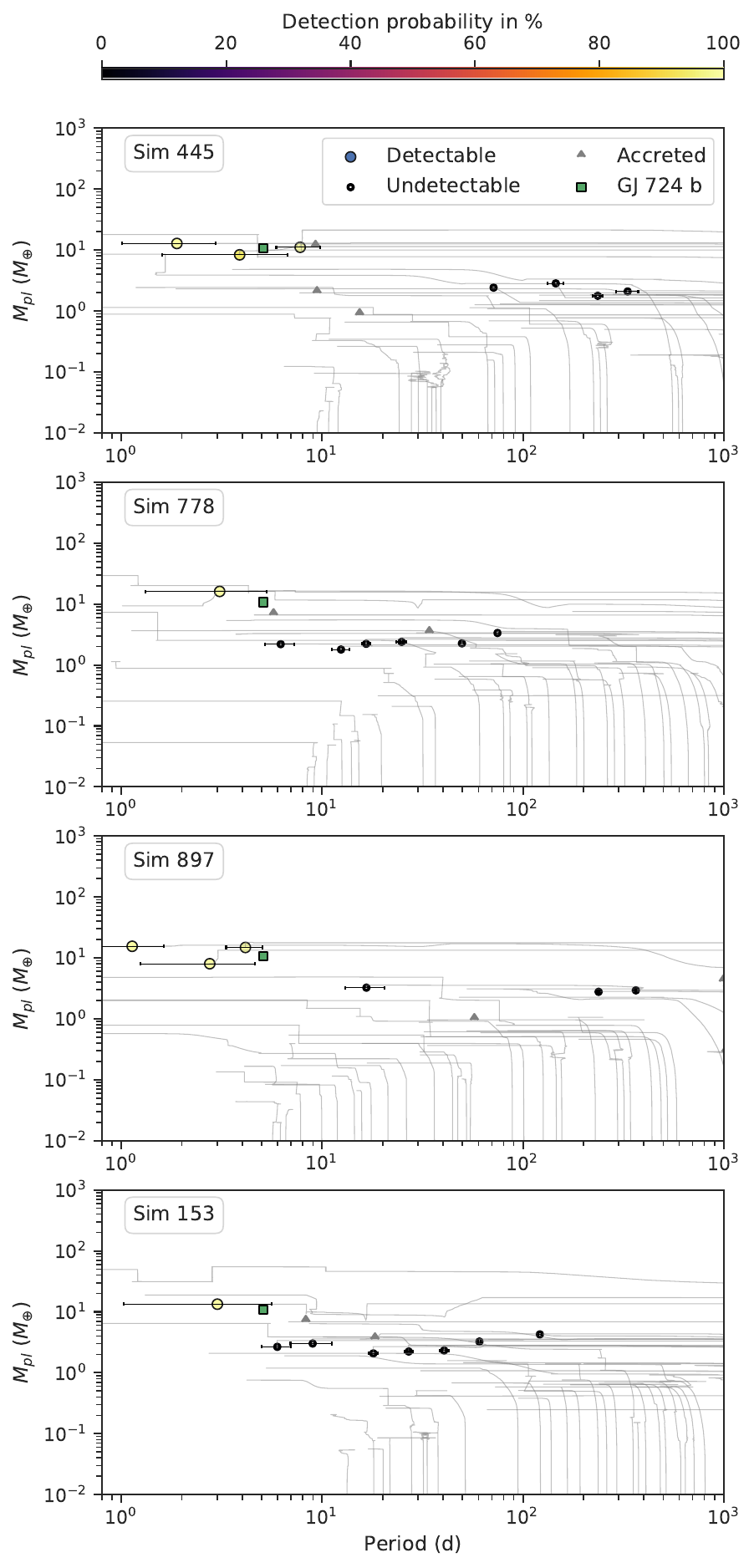}
        \caption{Planetary mass against orbital period of theoretically computed planetary systems. The four systems are selected from the 1000 synthetic systems of \citet{Burn2021} by filtering for observable planets with $e>0.4$ and maximum displacement of 0.5 times the orbital period and mass of GJ 724 b.  Planets with detection probabilities above 50\% are plotted with bigger markers and thinner edges. Thin gray lines show the past history of each planet and triangles show planets accreted by the planet closest to GJ 724 b in mass-period space; for simulation 445, this is the second planet. Horizontal black error-bars indicate the distance from apastron to periastron of the planets measured in periods of circular orbits. Two pathways leading to different numbers of observable planets can be distinguished with only the pathway with recent accretion of a massive counterpart and a unobservable exterior chain of planets leading to a single detectable planet.}
        \label{fig:synthetic_tracks}
\end{figure}

Despite the failure to match the exact properties of GJ~724\,b, it is insightful to analyse systems with eccentric planets. We selected four systems containing planets closest to GJ~724\,b in mass, period, and eccentricity space and show the full evolution of the systems in \autoref{fig:synthetic_tracks}. Two of the systems (Sim 445 and 897) have a tightly packed group of similar-mass planets, which can be ruled out by observations. Interestingly, the other two simulations (Sim 778 and 153), which only have one observable planet, show a very similar formation history. In both systems, while the disk was present, there were two planets with $\sim$10\,M$_\oplus$ locked in a mean-motion resonance close to the inner edge of the disk. However, over Gyr timescales, the innermost planet spirals into the star due to tidal interactions leaving only the outer planet which experiences weaker tides due to its lower mass, more compact structure, and larger orbital period. This is a promising pathway for systems like the one around GJ 724; however, more research into the exact dynamical evolution of this configuration is required.

Alternatively, close-in planets in eccentric orbits could undergo high-eccentricity tidal migration and circularisation \citep[e.g.][]{Rasio1996, Rice2012, Giacalone2017, Dong2021, Yu2021, Yu2022}, a process not included in the aforementioned models \citep{Emsenhuber2021}. Similarly to the synthetic planets discussed above, eccentric inner planets can form during the early stages of the planetary system evolution by interacting with massive counterparts \citep[e.g.][]{Weidenschilling1996, Chatterjee2008, Juric2008, Raymond2009, Carrera2019}. In order to check these two possibilities,  we computed an estimate of the circularisation timescale. We did this by following Eqs. 1 and 2 of \cite{Jackson2009}. 



Since we do not know the radius of the planet, we have to rely on an estimate based on its characteristics. We measured only the minimum planetary mass of GJ~724\,b, which is $10. 75^{+0.96}_{-0.87}\, M_\oplus$,  for which we can assume that it has around Neptune's mass. Following the mass-radius relation in \citealt{Luque2022}, we can expect a non-rocky planet. Using Neptune's density, we obtained a radius of $3.5$ Earth radii. We also adopt a tidal dissipation value ($Q$) for a possible Neptune-like planet of $Q_p=\num{1e5}$ \citep{Tittemore1990, Banfield1992, Correia2020} and $Q_\star=\num{1e5}$ for the star, as it is the most commonly used value \citep[e.g][]{Lin1996}. This results in a circularisation timescale of $1.0 - 2.0$ \si{\giga\year} for planet's radius of $3.5 - 4.0$ Earth radii. 
The stellar age is not known, but with a nearly solar metallicity, we can assume roughly  solar age or \SI{5}{\giga\year}. The circularisation timescale is therefore shorter than the age of the system, which means that to explain our observation the planet would have to have suffered from a recent perturbation event, since the planet would remain within the 1\,$\sigma$ uncertainty interval of the determined eccentricity for only about 
\SI{0.5}{\giga\year}. Alternatively, the eccentricity and semi-major axis can be evolved back in time by changing the sign of Eqs. 1 and 2 from \cite{Jackson2009}. With the same assumptions, \SI{5}{\giga\year} ago the planet would have had an eccentricity of 0.9, which means that it would have suffered from a scattering event at an early stage of the planetary system. This seems to be a more plausible explanation. Unfortunately, due to the uncertainty in the $Q$ values, we cannot give a more precise result as we only have an estimate of from the solar system.


We also considered the possibility of a Kozai effect in the system induced by another companion, as in the case of GJ~436\,b \citep{Bourrier2018}. The perturber could be another planet, a brown dwarf, or, under certain circumstances, a very low-mass star. As mentioned in \autoref{sec:properties}, GJ~724 does not seem to be part of a binary or multi-stellar system. Whereas for the planetary companion, we already discarded the possibility of a Jupiter-like planet in a Jupiter-like orbit with the detection map (\autoref{fig:GJ724_detection_map}). Nevertheless, a possible Kozai effect cannot be completely ruled out as an explanation for the high eccentricity of GJ~724\,b. In this sense, more data is desirable to search for possible companions that had not been detected with our dataset. Future studies based on dynamical interactions could also check if such an (undetected) companion can cause such high eccentricity, for which we strongly suggest future works on this system.


\subsection{GJ~3988}
The orbital fit of GJ~3988 system is unambiguous as it is consistent with a circular orbit. Based on the planet's minimum mass ($M_b\sin{i}= 3.69^{+0.42}_{-0.41}\, M_\oplus$), we expect GJ~3988\,b to be a super-Earth or mini-Neptune, as we do not know its radius and bulk density.  Due to its proximity to the star ($ a= 0.0405^{+0.0011}_{-0.0012} $ \si{\astronomicalunit}), this planet (if it indeed Neptune-like) is likely to undergo atmospheric mass loss due to the emission received from its host star, or to lose its atmosphere if is a rocky planet \citep[e.g.][]{Lammer2003, LecavelierdesEtangs2004,Lammer2009, Erkaev2016, Owen2016, Lehmer2017, Locci2019}.

\begin{figure}
    \centering
    \includegraphics[width=\hsize]{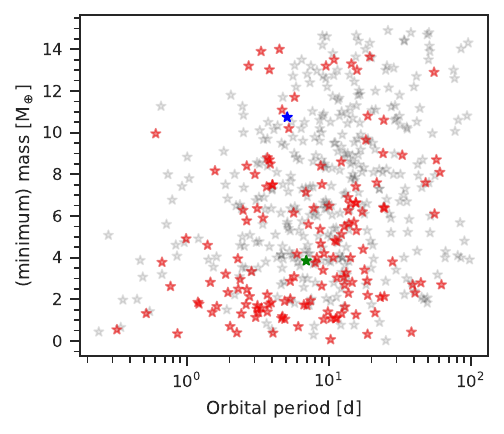}
    \caption{(Minimum) mass vs. orbital period plot of confirmed exoplanets from \href{https://exoplanetarchive.ipac.caltech.edu/}{NASA Exoplanet Archive} with orbital periods less than \SI{100}{\day} and minimum masses less than $15\, M_\oplus$. The red stars represents planets around M-dwarfs stars, whereas the gray colour depicts planets around other types of stars. The green star represents the position of GJ~3899\,b in the diagram. For completeness, we also include GJ~724\,b, represented as a blue star. }
    \label{fig:GJ3988_period_mass}
\end{figure}

There are other exoplanets with characteristics (orbital period and minimum mass) similar to those of GJ~3899\,b. For instance, G~264-012\,c \citep{Amado2021} is a terrestrial planet ($M_c\sin{i}= 3.75^{+0.48}_{-0.47}\,M_\oplus$ ) orbiting in a circular trajectory around an M4.0 dwarf. Its orbital period is \SI{8.5}{\day} and its semi-major axis is $0.02279 \pm 0.00061$ \si{\astronomicalunit}; these parameters are comparable to those of  GJ~3988\,b. Another example is  GJ~3138\,b \citep{Astudillo-Defru2017a}, orbiting in a nearly-circular orbit ($e=0.11^{+0.11}_{-0.07}$) around an M0 star. Its minimum mass ($M_b\sin{i}= 4.18^{+0.61}_{-0.59}\,M_\oplus$) is in the regime of super-Earths and mini-Neptunes.  It orbits its host star with a short period ($P=5.974 \pm 0.001$ \si{\day}, $a=0.057 \pm 0.001$ \si{\astronomicalunit}), similarly to GJ~3988\,b. Additionally, TOI-776\,b \citep{Luque2021}, is a transiting planet orbiting a bright M1 V star. Its mass is $M_b=4.0 \pm 0.9\,M_\oplus$ and its orbital period is \SI{8.25}{\day}. As the other examples, this planet orbits near to its host star ($a=0.0652 \pm 0.0014$\,au). In \autoref{fig:GJ3988_period_mass} we display a (minimum) mass against orbital period plot of small ($M <15 \, M_\oplus$) and close-in ($P <100$ \si{\day}) confirmed exoplanets, demonstrating that GJ~3988\,b represents a common type of planet around M dwarfs. 

\section{Summary}
\label{sec:conclusions}

We discovered GJ~724\,b and GJ~3988\,b using RV measurements from CARMENES, HARPS, and IRD. We measured the mean stellar rotation periods for both stars using photometric data, obtaining \SI{116 \pm 2}{\day} for GJ~3988, and \SI{57 \pm 1}{\day} for GJ~724, although we cannot completely exclude its second harmonic ($\sim 28$ \si{\day}) as a possible rotation period. We also examined  spectroscopic activity indicators and computed their {\tt GLS} periodograms to look for stellar activity signals. Both planetary signals are dominant in the RV periodograms and thus easy to detect. We analysed their stability and concluded that they are not caused by stellar activity. We tested different configurations for both systems and found that the best models with the strongest evidences corresponded to single planets.

The minimum masses of GJ~724\,b and GJ~3988\,b are $10.75^{+0.96}_{-0.87}\,M_\oplus$ and $3.69^{+0.42}_{-0.41}\,M_\oplus$, while their orbital periods are \SI{5.10}{\day} and \SI{6.94}{\day}. The separations from their host stars are less than \SI{0.05}{\astronomicalunit}. GJ~724\,b has a very eccentric orbit ($e=0.577^{+0.055}_{-0.052}  $), having the highest eccentricities of all single planets around M dwarfs. We discussed some possibilities for this configuration, such as another (undetectable) companion in the system, tidal in-spiral migration, perturbation or scattering event in the early stage of the system, or a Kozai effect. With the aim to constrain our current knowledge of planetary formation and architecture, we suggest to obtain more data to verify our measurement itself and to search for possible companions. We also encourage future dynamical interaction studies, which would also help to improve our understanding of the high eccentricity of GJ~724\,b. On the other hand,  GJ~3988\,b orbits around its host star in a circular trajectory, as do a fair number of planets around M dwarfs, making it a common planet around these type of stars.


\begin{acknowledgements}

This publication was based on observations collected under the CARMENES Legacy+ project.
  
CARMENES is an instrument at the Centro Astron\'omico Hispano en Andaluc\'ia (CAHA) at Calar Alto (Almer\'{\i}a, Spain), operated jointly by the Junta de Andaluc\'ia and the Instituto de Astrof\'isica de Andaluc\'ia (CSIC).
  
CARMENES was funded by the Max-Planck-Gesellschaft (MPG), 
  the Consejo Superior de Investigaciones Cient\'{\i}ficas (CSIC),
  the Ministerio de Econom\'ia y Competitividad (MINECO) and the European Regional Development Fund (ERDF) through projects FICTS-2011-02, ICTS-2017-07-CAHA-4, and CAHA16-CE-3978, 
  and the members of the CARMENES Consortium 
  (Max-Planck-Institut f\"ur Astronomie,
  Instituto de Astrof\'{\i}sica de Andaluc\'{\i}a,
  Landessternwarte K\"onigstuhl,
  Institut de Ci\`encies de l'Espai,
  Institut f\"ur Astrophysik G\"ottingen,
  Universidad Complutense de Madrid,
  Th\"uringer Landessternwarte Tautenburg,
  Instituto de Astrof\'{\i}sica de Canarias,
  Hamburger Sternwarte,
  Centro de Astrobiolog\'{\i}a and
  Centro Astron\'omico Hispano-Alem\'an), 
  with additional contributions by the MINECO, 
  the Deutsche Forschungsgemeinschaft (DFG) through the Major Research Instrumentation Programme and Research Unit FOR2544 ``Blue Planets around Red Stars'', 
  the Klaus Tschira Stiftung, 
  the states of Baden-W\"urttemberg and Niedersachsen, 
  and by the Junta de Andaluc\'{\i}a.

Data were partly collected with the 90\,cm telescope at the Observatorio de Sierra Nevada operated by the Instituto de Astrof\'\i fica de Andaluc\'\i a (CSIC). 

The authors wish to express their sincere thanks to all members of the Calar Alto and Sierra Nevada staff for their expert support of the instrument and telescope operation.
  
  
  We acknowledge financial support from the Agencia Estatal de Investigaci\'on (AEI/10.13039/501100011033) of the Ministerio de Ciencia e Innovaci\'on and the ERDF ``A way of making Europe'' through projects 
  PID2021-125627OB-C31,         
  PID2019-109522GB-C5[1:4],     
  PID2019-107061GB-C64,     
  PID2019-110689RB-100,     
  CEX2021-001131-S,         
and the Centre of Excellence ``Severo Ochoa'' and ``Mar\'ia de Maeztu'' awards to the Instituto de Astrof\'isica de Canarias (CEX2019-000920-S), Instituto de Astrof\'isica de Andaluc\'ia (SEV-2017-0709) and Institut de Ci\`encies de l'Espai (CEX2020-001058-M).

This work was also funded by the DFG through grant DR\,281/39-1, the Israel Science Foundation through grant No. 1404/22, the Generalitat de Catalunya/CERCA programme, the JSPS KAKENHI grant No.18H05442.

We recognise the support from Silvia Sabotta for creating the detection map of GJ~724.

We thank the anonymous referee for the comments that improved the quality of this manuscript.

\end{acknowledgements}

\bibliographystyle{aa}
\bibliography{mybib}

\appendix

{\onecolumn}

\section{Priors used in the models}

\begin{table*}[!ht]
    \centering
    \caption{Default priors used for the GP kernels.}
    \label{tab:priors_GPs}
    \begin{tabularx}{\hsize}{l c c X}
        \hline
        \hline
        \noalign{\smallskip}
        Parameter                                       & Prior                               & Unit                   & Description \\
        \noalign{\smallskip}
        \hline
        \noalign{\smallskip}
        \multicolumn{4}{c}{\textit{dSHO-GP kernel parameters}}          \\
        \noalign{\smallskip}
        $P_\text{GP}$                          & $\mathcal{U}$ / $\mathcal{N}$       & d                      & Period, uniform range for explorative fits; normal distribution centred on the stellar rotation period for RV and photometric fits, with at least three times the uncertainty of the period as the standard deviation of the normal distribution \\
        $\sigma_\text{GP}$  & $\mathcal{U}(0,50)$ / $\mathcal{J}(1,\num{1e6})$            & \si{\meter\per\second} / \si{ppm} & Standard deviation of the GP, uniform for RVs and log-uniform for photometry \\
        $f_\text{GP}$                               & $\mathcal{U}(0, 1)$             & \dots                  & Fractional amplitude of secondary mode \\
        $Q_{0, \text{GP}}$                          & $\mathcal{J}(0.1, \num{1e5})$        & \dots                  & Quality factor of secondary mode \\
        $dQ_\text{GP}$                              & $\mathcal{J}(0.1, \num{1e5})$        & \dots                  & Difference in quality factor between primary and secondary mode \\
        \noalign{\smallskip}
       \multicolumn{4}{c}{\textit{QP-GP kernel parameters}} \\
       \noalign{\smallskip}
        $P_\text{GP}$                           & $\mathcal{U}$ / $\mathcal{N}$       & d                      &  Period of the GP quasi-periodic component \\            
        $\sigma_\text{GP}$  &  $\mathcal{J}(1,\num{1e6})$            & \si{ppm} & Amplitude of the GP component \\
        $\Gamma_\text{GP}$  & $\mathcal{U}(0,10)$             & \dots & Amplitude of the GP sine-squared component \\
        $\alpha_\text{GP}$  & $\mathcal{U}(0,10)$            & \si{\day}$^{-2}$ & Inverse length-scale of the GP exponential component \\
        \noalign{\smallskip}
        \multicolumn{4}{c}{\textit{QPC-GP kernel parameters}} \\  
        \noalign{\smallskip} 
        $P_\text{GP}$                           & $\mathcal{U}$ / $\mathcal{N}$       & d                      &  Period of the GP quasi-periodic component \\            $\sigma_\text{GP}$  & $\mathcal{J}(1,\num{1e6})$            &  \si{ppm} &  Amplitude of the GP \\
        $h_1$  & $\mathcal{J}(1,\num{1e5})$ & \si{\meter\per\second} & Amplitude of the quasi-periodic component \\
        $h_2$  & $\mathcal{J}(1,\num{1e5})$ & \si{\meter\per\second} & Amplitude of the cosine component \\
        \noalign{\smallskip}
        \hline
    \end{tabularx}
    \tablefoot{The prior labels $\mathcal{U}$, $\mathcal{J}$, and $\mathcal{N}$ represent uniform, log-uniform, and normal distributions, respectively.}
\end{table*}

\begin{table*}[!ht]
    \centering
    \caption{Priors used for the RVs fist to the RVs of GJ~724.}
    \label{tab:priors_GJ724}
    \begin{tabular}{l c c l}
        \hline
        \hline
        \noalign{\smallskip}
        Parameter                                       & Prior                               & Unit                   & Description                                                     \\
        \noalign{\smallskip}
        \hline
        \noalign{\smallskip}
        \multicolumn{4}{c}{\textit{Planet parameters}}                                                                                                                                   \\
        \noalign{\smallskip}
        $P_\text{b}$                                    & $\mathcal{U}(4, 6)$            & d                      & Period                                                          \\
        $K_\text{b}$                                    & $\mathcal{U}(0.0, 50.0)$            & \si{\meter\per\second} & RV semi-amplitude                                               \\
        $t_\text{0, b}$ (BJD)                           & $\mathcal{U}(2457508.0, 2457514.0)$ & d                      & Time of periastron passage                                      \\
        $\sqrt{e_\text{b}}\sin \omega_\text{b}$         & $\mathcal{U}(-1, 1)$ / fixed(0)   & \dots                  & {Parameterisation} for $e$ and $\omega$.                        \\
        $\sqrt{e_\text{b}}\cos \omega_\text{b}$         & $\mathcal{U}(-1, 1)$  / fixed(0)    & \dots                  & {Parameterisation} for $e$ and $\omega$.                        \\
        
       \noalign{\smallskip}
       
        $P_\text{c}$                                    & $\mathcal{U}(1.5, 3.5)$            & d                      & Period                                                          \\
        $K_\text{c}$                                    & $\mathcal{U}(0.0, 50.0)$            & \si{\meter\per\second} & RV semi-amplitude                                               \\
        $t_\text{0, c}$ (BJD)                           & $\mathcal{U}(2457505.0, 2457508.5)$ & d                      & Time of periastron passage                                      \\
        $\sqrt{e_\text{c}}\sin \omega_\text{c}$         & fixed(0)   & \dots                  & {Parameterisation} for $e$ and $\omega$.                        \\
        $\sqrt{e_\text{c}}\cos \omega_\text{c}$         & fixed(0)    & \dots                  & {Parameterisation} for $e$ and $\omega$.                        \\
        \noalign{\smallskip}
        \multicolumn{4}{c}{\textit{GP parameters}}                                                                                                                                       \\
        \noalign{\smallskip}
        $P_\text{GP, rv}$ (d)                           & $\mathcal{N}(28.0, 5.0)$  / $\mathcal{N}(56.0, 5.0)$       & d                      & Period                                                          \\
        $\sigma_\text{GP, CARM-VIS}$ (\si{\meter\per\second}) & $\mathcal{U}(0.0, 50.0)$            & \si{\meter\per\second} & Amplitude of the GP component from CARMENES RVs    \\
        $\sigma_\text{GP, HARPS}$ (\si{\meter\per\second}) & $\mathcal{U}(0.0, 50.0)$            & \si{\meter\per\second} & Amplitude of the GP component from HARPS RVs    \\                           
        $f_\text{GP, rv}$                               & $\mathcal{U}(0.0, 1.0)$             & \dots                  & Fractional amplitude of secondary mode                          \\
        $Q_{0, \text{GP, rv}}$                          & $\mathcal{J}(0.1, \num{1e5})$        & \dots                  & Quality factor of secondary mode                                \\
        $dQ_\text{GP, rv}$                              & $\mathcal{J}(0.1, \num{1e5})$        & \dots                  & Difference in quality factor between primary and secondary mode \\

        \noalign{\smallskip}
        \multicolumn{4}{c}{\textit{Instrument parameters}}                                                                                                                               \\
        \noalign{\smallskip}
        $\gamma_\text{CARM-VIS}$                           & $\mathcal{U}(-10, 10)$              & \si{\meter\per\second} & RV zero point from CARMENES RVs                                                \\
        $\gamma_\text{HARPS}$                           & $\mathcal{U}(-10, 10)$              & \si{\meter\per\second} & RV zero point from HARMENES RVs                                                \\
        $\sigma_\text{CARM-VIS}$                        & $\mathcal{U}(0.0, 30)$              & \si{\meter\per\second} & A jitter added in quadrature from CARMENES RVs                                \\
        $\sigma_\text{HARPS}$                        & $\mathcal{U}(0.0, 30)$              & \si{\meter\per\second} & A jitter added in quadrature from HARPS RVs                                \\
        \hline
    \end{tabular}
    \tablefoot{The prior labels $\mathcal{U}$, $\mathcal{J}$, and $\mathcal{N}$ represent uniform, log-uniform, and normal distributions, respectively.}
\end{table*}

\begin{table*}[!ht]
    \centering
    \caption{Priors used for the RVs fits to the RVs of GJ~3988.}
    \label{tab:priors_GJ3988}
    \begin{tabular}{l c c l}
        \hline
        \hline
        \noalign{\smallskip}
        Parameter                                       & Prior                               & Unit                   & Description                                                     \\
        \noalign{\smallskip}
        \hline
        \noalign{\smallskip}
        \multicolumn{4}{c}{\textit{Planet parameters}}                                                                                                                                   \\
        \noalign{\smallskip}
        $P_\text{b}$                                    & $\mathcal{U}(0.0, 20.0)$            & d                      & Period                                                          \\
        $K_\text{b}$                                    & $\mathcal{U}(0.0, 50.0)$            & \si{\meter\per\second} & RV semi-amplitude                                               \\
        $t_\text{0, b}$ (BJD)                           & $\mathcal{U}(2457508.0, 2457512.0)$ & d                      & Time of periastron passage                                      \\
        $\sqrt{e_\text{b}}\sin \omega_\text{b}$         & $\mathcal{U}(-1, 1)$ / fixed(0)     & \dots                  & {Parameterisation} for $e$ and $\omega$.                        \\
        $\sqrt{e_\text{b}}\cos \omega_\text{b}$         & $\mathcal{U}(-1, 1)$ / fixed(0)     & \dots                  & {Parameterisation} for $e$ and $\omega$.                        \\
        
        $P_\text{c}$                                    & $\mathcal{U}(110,120)$            & d                      & Period                                                          \\
        $K_\text{c}$                                    & $\mathcal{U}(0.0, 50.0)$            & \si{\meter\per\second} & RV semi-amplitude                                               \\
        $t_\text{0, c}$ (BJD)                           & $\mathcal{U}(2457500,2457700)$ & d                      & Time of periastron passage                                      \\
        $\sqrt{e_\text{c}}\sin \omega_\text{b}$         & $\mathcal{U}(-1, 1)$ / fixed(0)     & \dots                  & {Parameterisation} for $e$ and $\omega$.                        \\
        $\sqrt{e_\text{c}}\cos \omega_\text{b}$         & $\mathcal{U}(-1, 1)$ / fixed(0)     & \dots                  & {Parameterisation} for $e$ and $\omega$.                        \\

        \noalign{\smallskip}
        \multicolumn{4}{c}{\textit{GP parameters}}                                                                                                                                       \\
        \noalign{\smallskip}
        $P_\text{GP, rv}$ (d)                           & $\mathcal{N}(116.0, 10.0)$          & d                      & Period                                                          \\
        $\sigma_\text{GP, CARM-VIS}$ (\si{\meter\per\second}) & $\mathcal{U}(0.0, 50.0)$            & \si{\meter\per\second} & Standard deviation                                              \\
        $\sigma_\text{GP, IRD}$ (\si{\meter\per\second}) & $\mathcal{U}(0.0, 50.0)$            & \si{\meter\per\second} & Standard deviation                                              \\
        $f_\text{GP, rv}$                               & $\mathcal{U}(0.0, 1.0)$             & \dots                  & Fractional amplitude of secondary mode                          \\
        $Q_{0, \text{GP, rv}}$                          & $\mathcal{J}(0.1, \num{1e5})$        & \dots                  & Quality factor of secondary mode                                \\
        $dQ_\text{GP, rv}$                              & $\mathcal{J}(0.1, \num{1e5})$        & \dots                  & Difference in quality factor between primary and secondary mode \\

        \noalign{\smallskip}
        \multicolumn{4}{c}{\textit{Instrument parameters}}                                                                                                                               \\
        \noalign{\smallskip}
        $\gamma_\text{CARM-VIS}$                           & $\mathcal{U}(-10, 10)$              & \si{\meter\per\second} & RV zero point                                                   \\
        $\sigma_\text{CARM-VIS}$                        & $\mathcal{U}(0.0, 30)$              & \si{\meter\per\second} & A jitter added in quadrature                                    \\
        $\gamma_\text{IRD}$                           & $\mathcal{U}(-10, 10)$              & \si{\meter\per\second} & RV zero point                                                   \\
        $\sigma_\text{IRD}$                        & $\mathcal{U}(0.0, 30)$              & \si{\meter\per\second} & A jitter added in quadrature                                    \\
        \noalign{\smallskip}
        \hline
    \end{tabular}
    \tablefoot{The prior labels $\mathcal{U}$, $\mathcal{J}$, and $\mathcal{N}$ represent uniform, log-uniform, and normal distributions, respectively.}
\end{table*}

\clearpage
\newpage
\section{Plots from the best models}

\begin{figure}[!ht]
\centering
\includegraphics{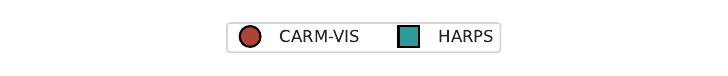}
\includegraphics{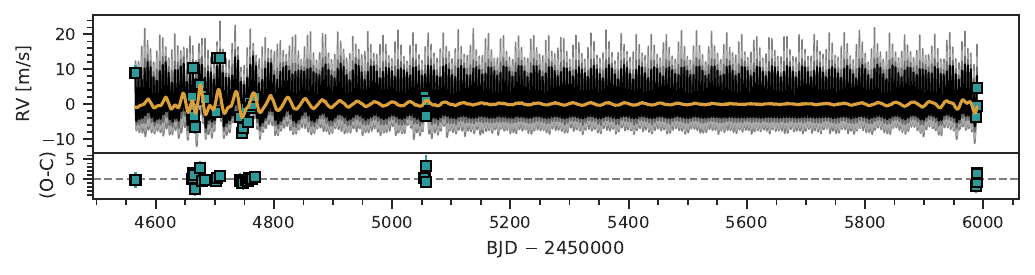}
\includegraphics{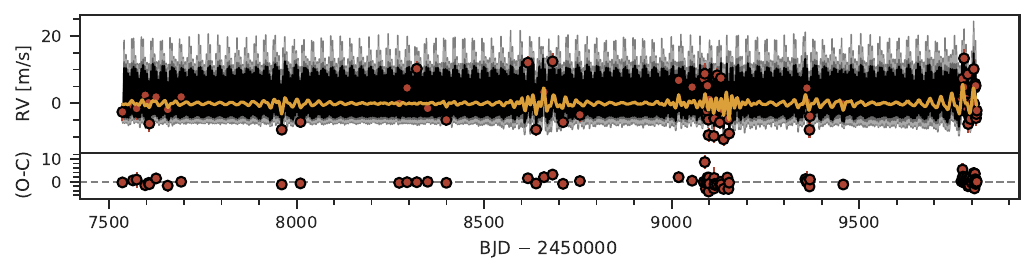}
\caption{RVs over time for the best model ($\text{1P}_\text{(5\,d-ecc)} + \text{dSHO-GP}_\text{28\,d}$) fitted to the HARPS (top) and CARMENES (bottom) RVs of GJ~724. The GP component of the model is plotted as the orange solid line. The black lines show the median of \num{10000} samples from the posterior and the grey shaded areas denote the \SI{68}{\percent}, \SI{95}{\percent,} and \SI{99}{\percent} confidence intervals, respectively. Instrumental RV offsets were subtracted from the measurements and the model, and the error bars include the jitter added in quadrature. The residuals after subtracting the median model are shown in the lower panel. }
\label{fig:rvs_over_time_GJ724}
\end{figure}

\begin{figure}[!ht]
    \centering
    \includegraphics[width=1.0\hsize]{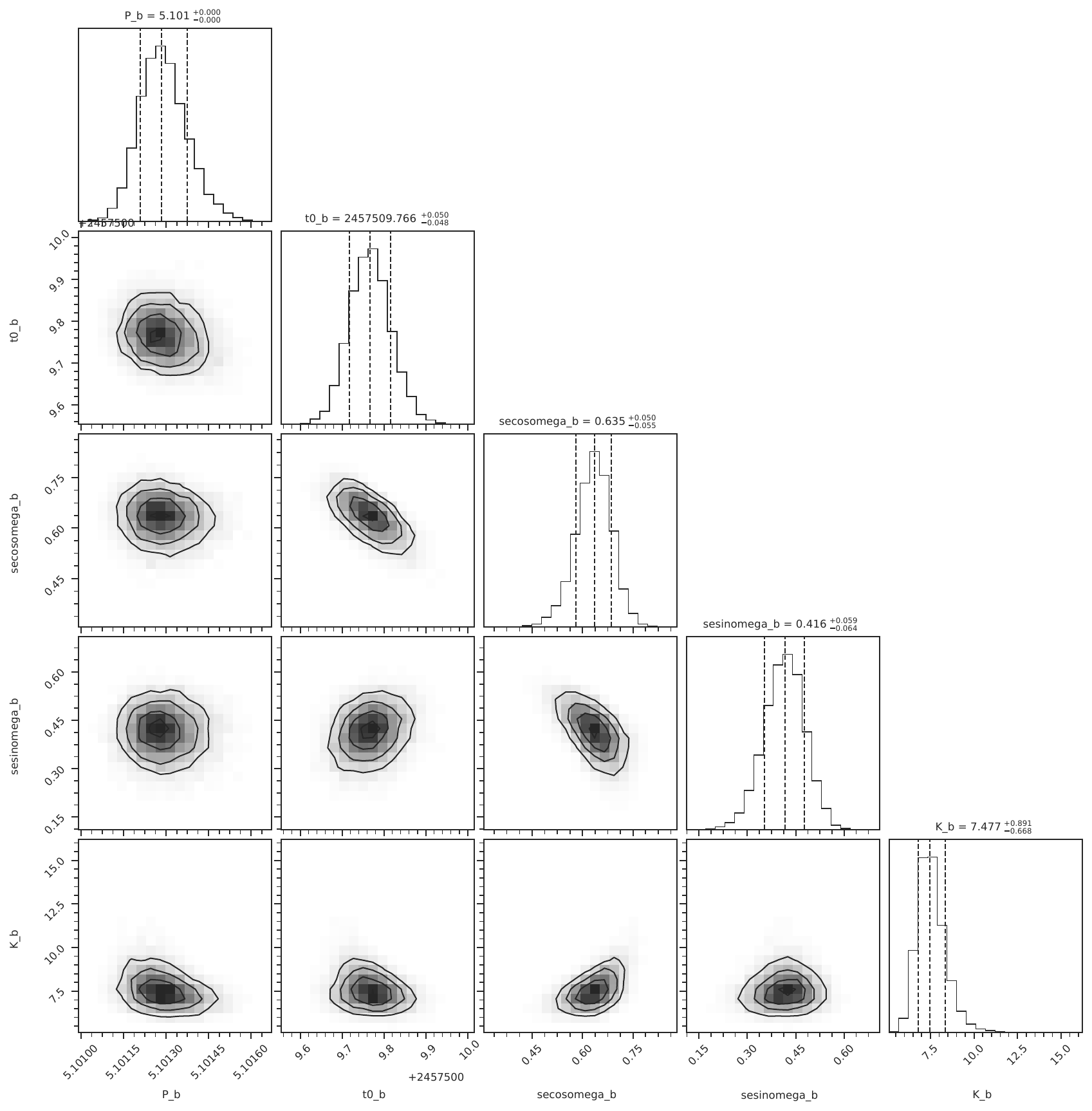} 
    \caption{Corner plot of the planetary parameters for GJ~724\,b from the best model ($\text{1P}_\text{(5\,d-ecc)} + \text{dSHO-GP}_\text{28\,d}$).}
    \label{fig:corner-planet-GJ724}
\end{figure}

\begin{figure}[!ht]
    \centering
    \includegraphics[width=1.0\hsize]{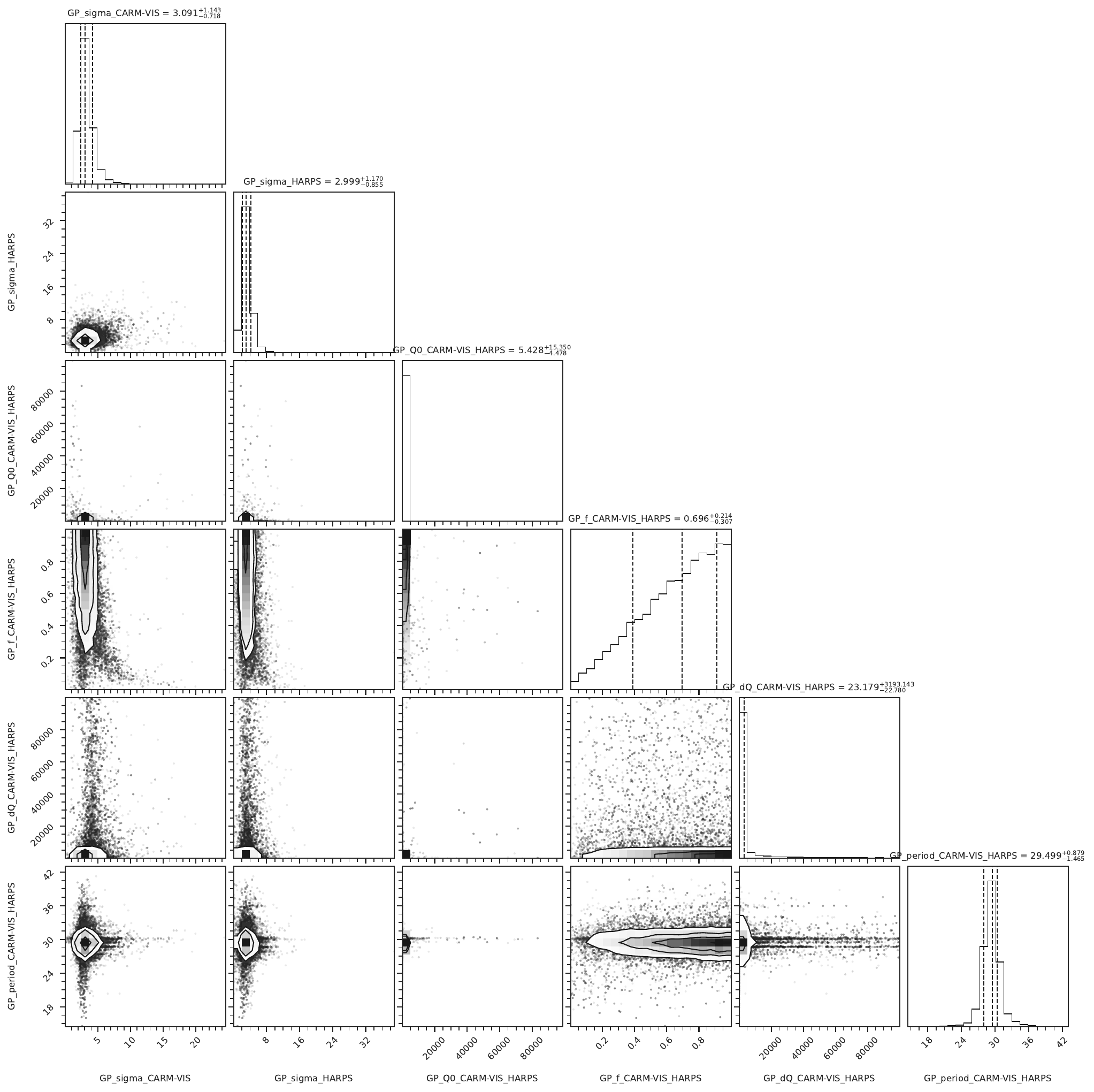} 
    \caption{Corner plot of the GP parameters from the best model of GJ~724 ($\text{1P}_\text{(5\,d-ecc)} + \text{dSHO-GP}_\text{28\,d}$).}
    \label{fig:corner-planet-GJ724-GP}
\end{figure}

\begin{figure}[!ht]
\centering
\includegraphics{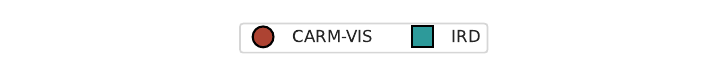} \\
\includegraphics{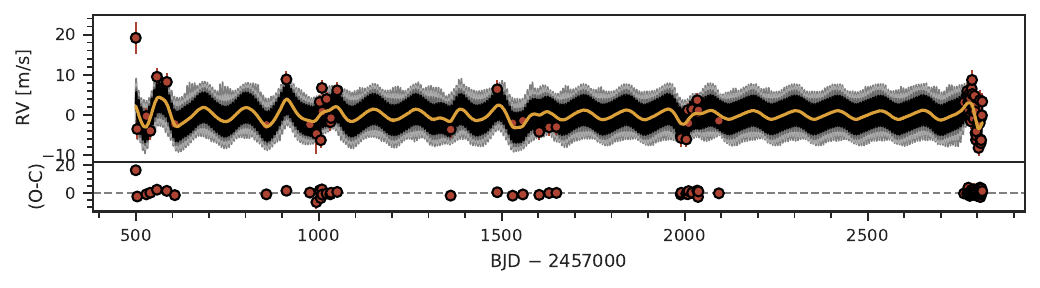}
\includegraphics{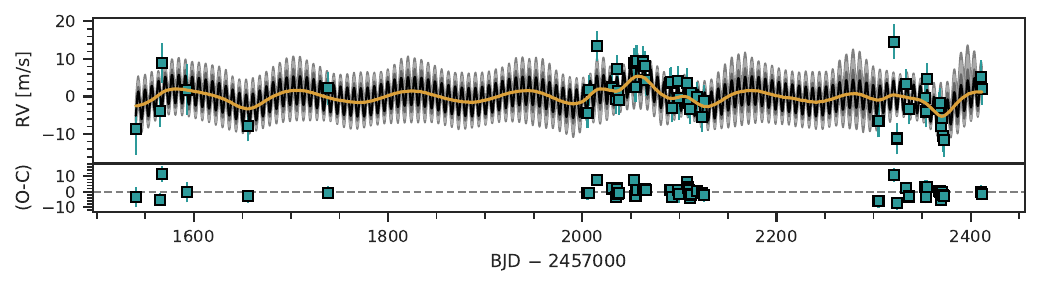}
\caption{Same as \autoref{fig:rvs_over_time_GJ724} but for the best model for GJ~3988 ($\text{1P}_\text{(7\,d-circ)} + \text{dSHO-GP}_\text{116\,d}$) fitted to the CARMENES (top) and IRD (bottom) RVs. }
\label{fig:rvs_over_time_GJ3899}
\end{figure}

\begin{figure}[!ht]
    \centering
    \includegraphics[width=0.7\hsize]{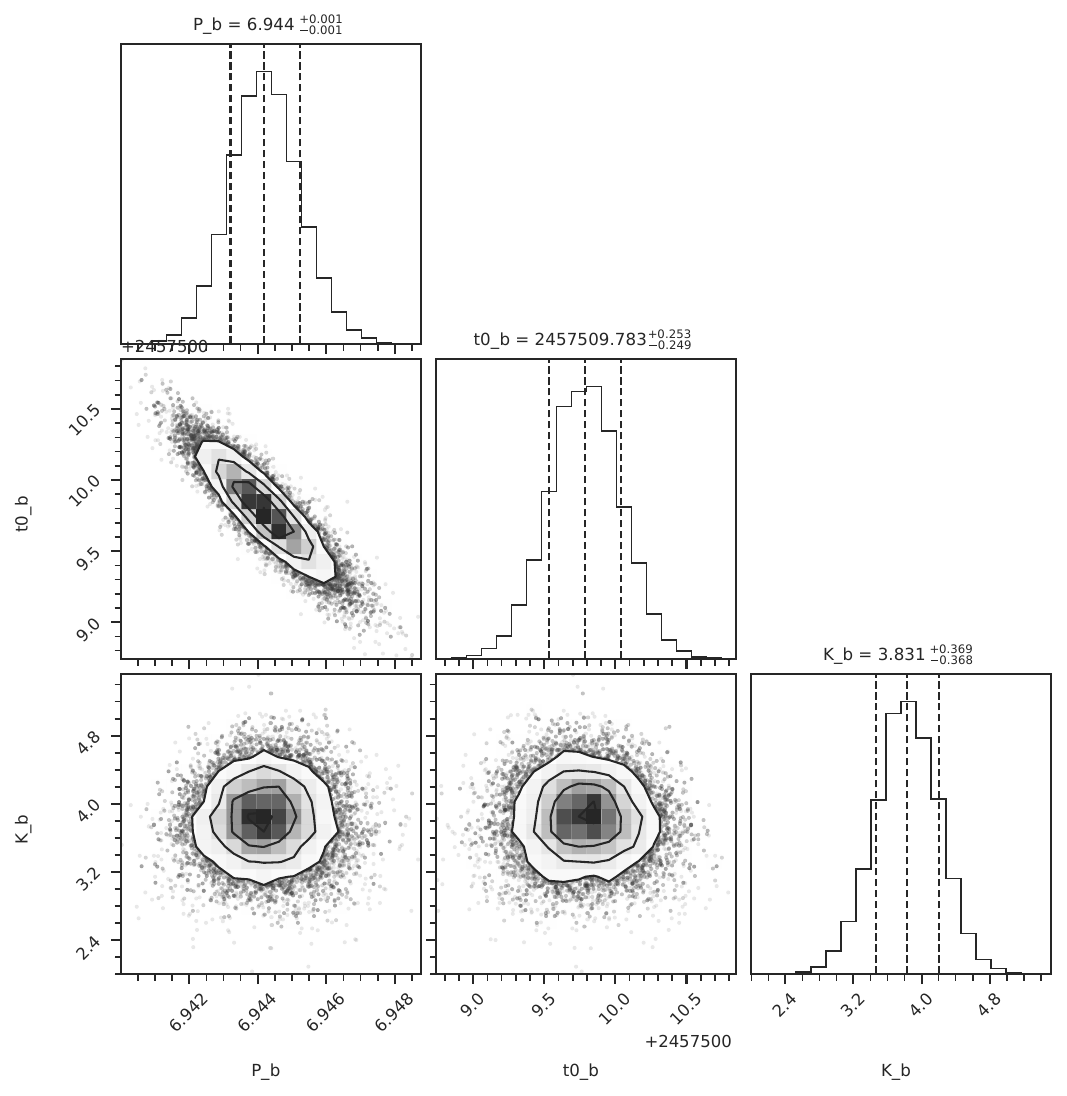} 
    \caption{Corner plot of the planetary parameters for GJ~3988\,b from the best model correspond to one planet ($\text{1P}_\text{(7\,d-circ)} + \text{dSHO-GP}_\text{116\,d}$).}
    \label{fig:corner-planet-GJ3988}
\end{figure}

\begin{figure}[!ht]
    \centering
    \includegraphics[width=1.0\hsize]{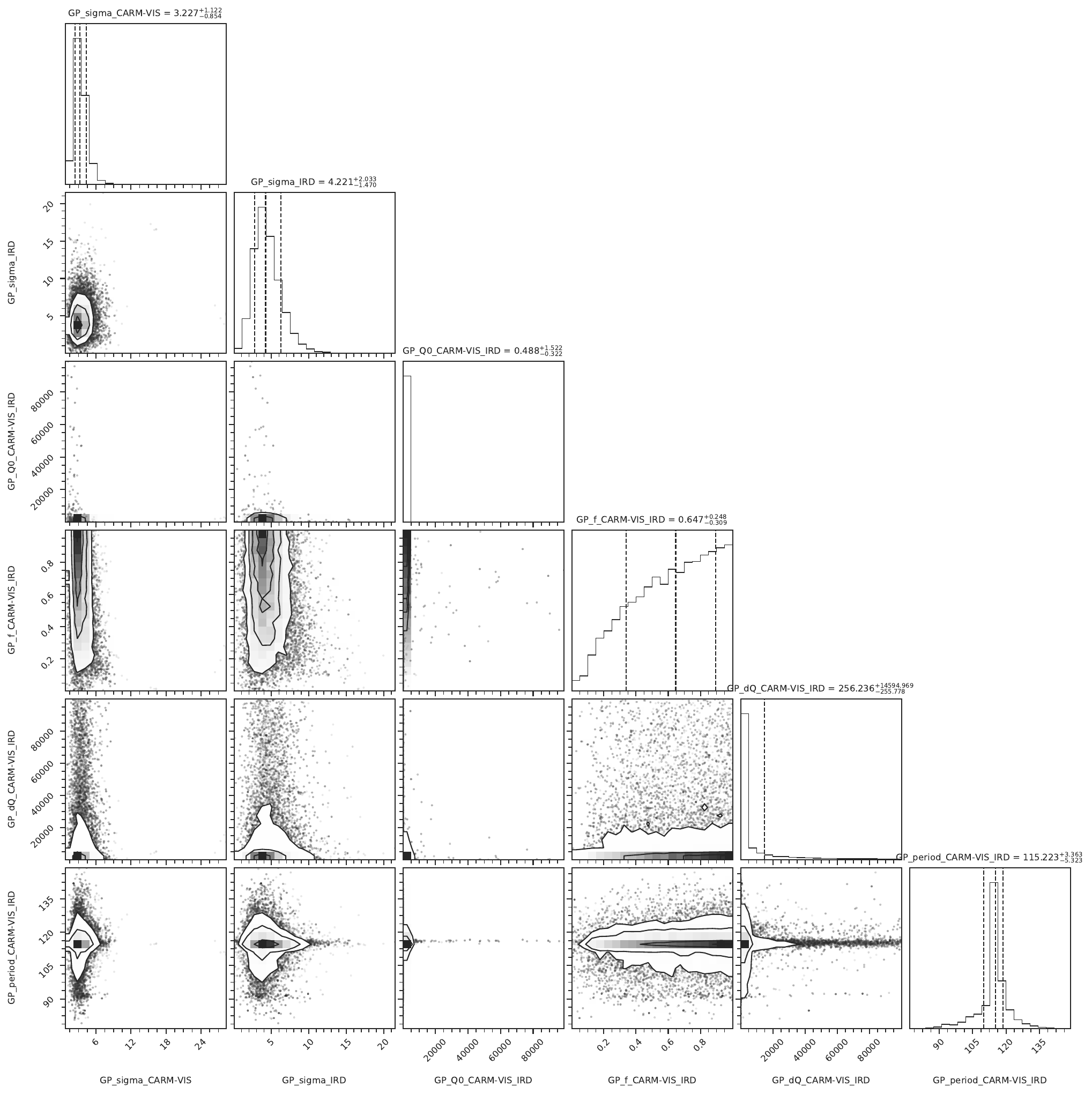} 
    \caption{Corner plot of the GP parameters from the best model of GJ~3988 ($\text{1P}_\text{(7\,d-circ)} + \text{dSHO-GP}_\text{116\,d}$).}
    \label{fig:corner-planet-GJ3988-GP}
\end{figure}

\clearpage
\newpage

\section{Comparison of posterior values}

\begin{figure}[!ht]
    \centering
    \includegraphics[width=0.48\hsize]{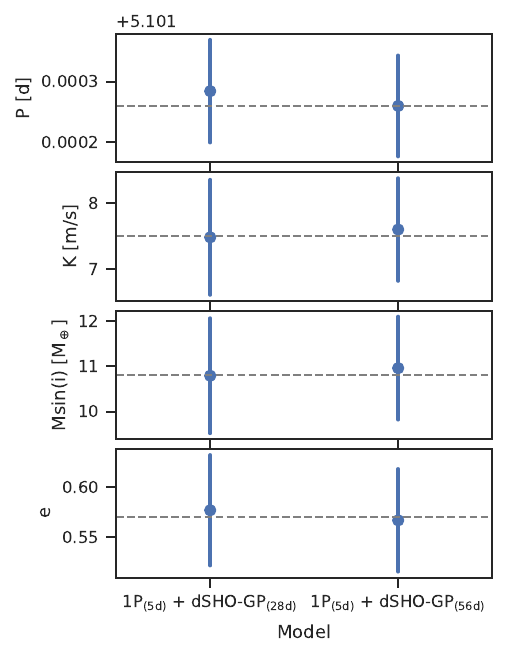} 
    \caption{Comparison of the posterior values of the main planetary parameters from two different models for GJ~724: one eccentric planet and a GP centred around  $P_\text{rot/2}=$ \SI{28}{\day} (left) and around  $P_\text{rot}=$ \SI{56}{\day} (right).}
    \label{fig:model-comp-p1}
\end{figure}

\end{document}